\documentclass[usenatbib,iop,numberedappendix]{aeb_emulateapj_2010}
\usepackage{amsmath}
\usepackage{amssymb}
\usepackage{xspace}
\usepackage[normalem]{ulem}

\usepackage{natbib}

\usepackage[usenames,dvips]{color}

\def\s{{\rm s}}

\def\yr{{\rm yr}} 
\def\Gyr{{\rm G}\yr}

\def\m{{\rm m}}

\def\cm{{\rm c}\m} 
 
\def\pc{{\rm pc}} 
\def\kpc{{\rm k}\pc}

\def\eV{{\rm eV}} 
\def\keV{{\rm k}\eV} 
 
\def\GeV{{\rm G}\eV} 
\def\TeV{{\rm T}\eV}

\def\erg{{\rm erg}} 
 
\def\K{{\rm K}}

\def\G{{\rm G}}

\def\del#1{{}}

\newcommand{\rmn}{\mathrm}
\newcommand{\e}{{\rm e}}

\def\vel{\upsilon}

\def\Lya{Ly$\alpha$\xspace}

\def\Fermi{{\em Fermi}\xspace}

\begin{document}

\title{
The Cosmological Impact of Luminous TeV Blazars III:\\
Implications for Galaxy Clusters and the Formation of Dwarf Galaxies}

\author{
Christoph Pfrommer\altaffilmark{1,2},
Philip Chang\altaffilmark{2,3},
and
Avery E.~Broderick\altaffilmark{2,4,5}
}
\altaffiltext{1}{Heidelberg Institute for Theoretical Studies, Schloss-Wolfsbrunnenweg 35, D-69118 Heidelberg, Germany; christoph.pfrommer@h-its.org}
\altaffiltext{2}{Canadian Institute for Theoretical Astrophysics, 60 St.~George Street, Toronto, ON M5S 3H8, Canada; aeb@cita.utoronto.ca, pchang@cita.utoronto.ca}
\altaffiltext{3}{Department of Physics, University of Wisconsin-Milwaukee, 1900 E. Kenwood Boulevard, Milwaukee, WI 53211, USA}
\altaffiltext{4}{Perimeter Institute for Theoretical Physics, 31 Caroline Street North, Waterloo, ON, N2L 2Y5, Canada}
\altaffiltext{5}{Department of Physics and Astronomy, University of Waterloo, 200 University Avenue West, Waterloo, ON, N2L 3G1, Canada}

\shorttitle{The Cosmological Impact of Blazar TeV Emission III}
\shortauthors{Pfrommer, Chang, \& Broderick}

\begin{abstract}
  A subset of blazars are powerful TeV emitters, dominating the extragalactic
  component of the very high energy gamma-ray universe ($E\gtrsim100\,\GeV$).
  These TeV gamma rays generate ultra-relativistic electron-positron pairs via
  pair production with the extragalactic background light.  While it has
  generally been assumed that the kinetic energy of these pairs cascade to GeV
  gamma rays via inverse Compton scattering, we have argued in \citet[][Paper I
  in this series]{BCP} that plasma beam instabilities are capable of dissipating
  the pairs' energy locally on timescales short in comparison to the
  inverse-Compton cooling time, heating the intergalactic medium (IGM) with a
  rate that is independent of density.  This dramatically increases the entropy
  of the IGM after redshift $z\sim2$, with a number of important implications
  for structure formation: (1) this suggests a scenario for the origin of the
  cool core (CC)/non-cool core (NCC) bimodality in galaxy clusters and groups.
  Early-forming galaxy groups are unaffected because they can efficiently
  radiate the additional entropy, developing a CC.  However, late-forming groups
  do not have sufficient time to cool before the entropy is gravitationally
  reprocessed through successive mergers---counteracting cooling and potentially
  raising the core entropy further. This may result in a population of X-ray dim
  groups/clusters, consistent with X-ray stacking analyses of optically selected
  samples.  Hence blazar heating works different than feedback by active
  galactic nuclei, which we show can balance radiative cooling but is unable to
  transform CC into NCC clusters on the buoyancy timescale due to the weak
  coupling between the mechanical energy to the cluster gas. (2) We predict a
  suppression of the Sunyaev-Zel'dovich (SZ) power spectrum template on angular
  scales smaller than $5\arcmin$ due to the globally reduced central pressure of
  groups and clusters forming after $z\sim1$.  This allows for a larger rms
  amplitude of the density power spectrum, $\sigma_8$, and may reconcile
  SZ-inferred values with those by other cosmological probes even after allowing
  for a contribution due to patchy reionization. (3) Our redshift dependent
  entropy floor increases the characteristic halo mass below which dwarf
  galaxies cannot form by a factor of approximately 10 (50) at mean density (in
  voids) over that found in models that include photoionization alone. This
  prevents the formation of late-forming dwarf galaxies ($z\lesssim2$) with
  masses ranging from $10^{10}$ to $10^{11}\,\rmn{M}_\sun$ for redshifts
  $z\sim2$ to 0, respectively. This may help resolve the ``missing satellites
  problem'' in the Milky Way of the low observed abundances of dwarf satellites
  compared to cold dark matter simulations and may bring the observed early star
  formation histories into agreement with galaxy formation models. At the same
  time, it explains the ``void phenomenon'' by suppressing the formation of
  galaxies within existing dwarf halos of masses $<
  3\times10^{10}\,\rmn{M}_\sun$ with a maximum circular velocity
  $<60~\rmn{km~s}^{-1}$ for $z\lesssim2$; hence reconciling the number of dwarfs
  in low-density regions in simulations and the paucity of those in
  observations.
\end{abstract}


\section{Introduction} \label{III}

Extragalactic relativistic jets are powered by accreting super-massive black
holes (or in general the engines of active galactic nuclei, AGNs) and are able
to carry an enormous amount of power out to cosmological distances. Blazars are
a subclass of AGNs where the jet opening angle of typically $\sim10^\circ$
contains our line-of-sight, allowing us to detect the Doppler-boosted radiation.
Blazars are the dominant extragalactic source class in the TeV sky with
currently 36 known objects out of 46 extragalactic sources \citep[of the
remaining 10, 4 are radio galaxies, 2 are starburst galaxies, and 4 are not yet
identified; for a review, see][]{Hinton+2009}.\footnote{For an
  up-to-date list/visualization of the extragalactic TeV sky, see\\
  http://www.mppmu.mpg.de/$\sim$rwagner/sources/ or\\
  http://tevcat.uchicago.edu/.}  Recent observations by the {\em Fermi} Space
Telescope and ground based imaging atmospheric Cherenkov telescopes (H.E.S.S.,
MAGIC, and VERITAS)\footnote{High Energy Stereoscopic System, Major Atmospheric
  Gamma Imaging Cerenkov Telescope, and Very Energetic Radiation Imaging Telescope
  Array System.}  demonstrated that most of the electromagnetic power is emitted
in the gamma-ray band. The most extreme blazars achieve energies of up to 10
TeV, giving rise to the class of high-energy peaked BL Lac objects (HBL) while
the somewhat less efficient accelerators in intermediate-energy peaked BL Lac
objects (IBL) are, in some cases, also able to reach energies beyond 100
GeV. The emission mechanism is thought to be inverse Compton scattering of
ultra-relativistic electrons in the jet giving rise to power-law energy spectra
that increase as a function of energy and peak at the maximum energy that the
accelerating process of the radiating relativistic electrons is able to deliver.
The universe is not transparent to very-high energy gamma-ray radiation (VHEGR;
$E\gtrsim 100~\GeV$), i.e., a beam of these energetic photons will necessarily
produce electron and positron pairs off of the extragalactic background light
(EBL), with typical mean free paths of VHEGRs ranging from 30~Mpc to 1~Gpc
depending upon gamma-ray energy and source redshift.  The pairs produced by
VHEGR radiation have typical Lorentz factors of $10^5-10^7$.

There are only two possible ultimate destinations for the kinetic energy of the
pairs: {\em the energy can either be channeled into lower energy (GeV) gamma-ray
  radiation (to which the universe is transparent), or heat the ambient medium
  with a partitioning factor that depends on the relative rates of the
  processes.} The first process has generally been assumed to dominate the
manner in which these pairs lose energy, almost exclusively through inverse
Compton scattering the cosmic microwave background (CMB) and EBL on a typical
mean free path of $(10-100)~\kpc$ today.  When the up-scattered gamma-ray is
itself a VHEGR the process repeats, creating a second generation of pairs and
up-scattering additional photons.  The result is an inverse Compton cascade
depositing the energy of the original VHEGR in gamma rays with energies
$\lesssim100~\GeV$.

There are, however, problems with this picture.  First, the expected inverse
Compton bump has not been seen in the spectra of luminous blazars around
$10\,\GeV$. This could imply the existence of intergalactic magnetic fields that
deflect the pairs out of our line-of-sight, hence reducing the inverse Compton
emission
\citep{Nero-Vovk:10,Tave_etal:10a,Tave_etal:10b,Derm_etal:10,Tayl-Vovk-Nero:11,Dola_etal:11,Taka_etal:11,Vovk+12}.
Typical lower limits for magnetic fields would then range from $10^{-19}\,\G$ to
$10^{-15}\,\G$, depending on the assumed duty cycle of blazars and are dominated
by void regions, which dominate a typical line-of-sight. The values at the upper
end are of astrophysical interest in the context of the formation of galactic
fields\footnote{After adiabatic contraction and a handful of windings, nG field strengths
  can be produced from an intergalactic magnetic field of
  $\sim10^{-15}\,\G$.}.  Second, the spectral shape of the unresolved
extragalactic gamma-ray background (EGRB), that has been measured by \Fermi,
exhibits a steep power-law at energies below $100\,\GeV$ \citep{Fermi_EGRB2010}.
If blazars contribute substantially to the EGRB, we would expect to see a
flattening toward high energies, due to the inverse Compton cascades, in
conflict with the \Fermi EGRB.  Traditionally, the EGRB is then used to
constrain the evolution of the luminosity density of VHEGR sources \citep[see,
e.g.,][]{Naru-Tota:06,Knei-Mann:08,Inou-Tota:09,Vent:10}.  Generally, it has
been found that these cannot have exhibited the dramatic rise in numbers by
$z\sim1$--$2$ seen in the quasar distribution.  That is, the co-moving number of
blazars must have remained essentially fixed, at odds with both the large-scale
mass assembly history of the universe, e.g., star formation history, and with
the luminosity history of similarly accreting systems, e.g., the quasar
luminosity density.

In our first companion paper of this series \citep[][hereafter Paper I]{BCP}, we
argued that instead of initiating an inverse Compton cascade, the pairs
dissipate their kinetic energy locally, heating the intergalactic medium (IGM).
We identified a process that operates on a timescale fast in comparison to
inverse Compton cooling, dominating the latter for luminous TeV blazars,
i.e., for HBL and IBL blazars with an equivalent isotropic luminosity of
$L\gtrsim 10^{42}\,\erg\,\s^{-1}$ above $100\,\GeV$.  Despite its dilute nature,
the VHEGR-generated beam of ultra-relativistic pairs, propagating through the
IGM, is susceptible to plasma beam instabilities.  While the commonly discussed
Weibel and two-stream instabilities are strongly suppressed by finite beam
temperatures, these are special cases of a general filamentary ``oblique''
instability which is far more virulent, and strongly insensitive to finite
temperature effects
\citep{Bret-Firp-Deut:04,Bret-Firp-Deut:05,Bret:09,Bret-Grem-Diec:10,Lemo-Pell:10}.

If these instabilities saturate at a rate that is comparable to their linear
growth rate, the beam kinetic energy is directly transferred to electrostatic
modes which rapidly dissipate locally, heating the IGM.  If this scenario or an
analogous, similarly efficient mechanism operates in practice, it necessarily
suppresses the inverse Compton cascades and naturally explains the absence of an
inverse Compton bump in the TeV blazar spectra without invoking an intergalactic
magnetic field. At the same time, it allows for a redshift evolution of TeV blazars
that is identical to that of quasars without overproducing the EGRB. In fact,
for plausible parameters of TeV blazar spectra, it is possible to explain the
high-energy part of the EGRB (Paper I) without the need to appeal to exotic phenomena
\citep[e.g., dark matter annihilation,][]{Cavadini+2011}.

By dissipating the pairs' energy into the IGM, plasma instabilities (or similar
processes) provide a novel mechanism for heating the universe.  Integrating over
the energy flux per mean free path of all known TeV blazars yields a luminosity
density, or equivalently a local heating rate, that dominates that of
photoheating by more than an order of magnitude at the present epoch, after
accounting for incompleteness corrections \citep[][hereafter Paper II]{CBP}. We
have demonstrated that the local TeV blazar luminosity function is consistent
with a scaled version of the quasar luminosity function \citep{Hopkins+07}, thus
the conservative assumption is that they evolve similarly, presumably due to the
same underlying accretion physics.  With this assumption, we showed in Paper II
that for the redshifts at which blazar heating is likely to be important,
$z\lesssim 3.5$, the heating rate will be relatively uniform throughout space.
Between $z\sim 3.5$ and 6 it may experience order 50\% fluctuations, and by
$z\gtrsim 6$, it will exhibit significant stochasticity with order unity
deviations.

This heating differs from other feedback prescriptions in an important way:
since the number density of EBL photons and that of TeV blazars are nearly
homogeneously distributed on cosmological scales, so is the resulting pair
density. Hence, the implied heating rate is also homogeneous, i.e., the
volumetric blazar heating rate is uniform and independent of IGM
density\footnote{The term ``blazar heating'' exclusively denotes the dissipation
  of gamma-ray luminosity of HBL and IBL blazars between $100\,\GeV$ and
  $10\,\TeV$ with an equivalent isotropic luminosity of $L\gtrsim
  10^{42}\,\erg\,\s^{-1}$ (see Paper II for details).}. Due to the large mean
free path of TeV-photons, which is much larger than the turn-around radius of
any virialized structures, the heating process is not expected to be dominated
by contributions in highly biased regions at late times for $z \lesssim
3.5$. (At early times, when blazar heating exhibits considerable fluctuations,
clustering bias could substantially modify the phenomenology of the heating
mechanism.)  The effect of a uniform heating rate is that the energy deposited
per baryon is substantially larger in more tenuous regions of the universe.  As
a result, underdense regions experience a larger temperature increase, producing
an inverted temperature-density relation in voids, asymptotically approaching
$T\propto \rho_g^{-1}$. Generally, we found in Paper II that without any fine
tuning it is possible to reproduce the inverted temperature-density relation at
$z=2-3$ inferred by high-redshift \Lya studies \citep{Bolton+08,Viel+09}, while
simultaneously satisfying the temperature constraints at $z=2$ \citep[e.g.,
those by][]{Lidz+10} and leaving the local \Lya forest unaffected.

In a follow-up paper by \citet{Puchwein+2011}, we used hydrodynamic simulations
of cosmological structure formation to explicitly demonstrate that blazar
heating provides not only an excellent description of the one- and two-point
statistics of high-redshift \Lya forest spectra but also the line width and
column density distribution.  This detailed agreement includes reproducing the
observed mean transmission, and is achieved using the most recent estimate of
the evolution of the photoionizing background without any tuning.  These
successes are due specifically to the salient properties of blazar heating, in
particular its excess energy injection into the low density IGM and its
continuous nature.

In this work, we are interested in the impact of such an important heating
mechanism on cosmological structure formation. While we propose a
well-motivated, physical heating mechanism that uses a certain type of plasma
instability, our conclusions concerning the thermal history of the IGM, the \Lya
forest, as well as structure formation remain robust in that they only rely on a
mechanism that dissipates the energy of TeV blazars independently of the density
and employs a redshift evolution similar to that of the quasar luminosity
function. To study the impact on structure formation, we turn to the evolution
of the entropy because in the absence of radiative cooling, entropy is conserved
upon adiabatic compression and can only be increased through dissipation of
gravitational energy in structure formation shocks or heating due to
photo-ionization. Hence, entropy is a unique thermodynamic variable with which
to learn about the impact of non-gravitational feedback processes.  We identify
two different classes of objects where the time variable minimum entropy, or
entropy floor, of the IGM induced by blazar heating might dramatically change
our present picture of structure formation: the structure of galaxy groups and
clusters, and the formation of dwarf galaxies.  Before addressing the
consequences of blazar heating for each, we introduce each separately,
highlighting the relevant outstanding problems.

\subsection{The entropy problem in galaxy groups}
\label{sec:intro_groups}

In the absence of non-gravitational energy injection, the X-ray luminosity of
clusters ($L_x$) is expected to exhibit a self-similar scaling with intracluster
medium (ICM) temperature ($T$) as $L_{x}\propto T^2$ from purely gravitational
and shock dynamics \citep{Kaiser86,Evrard+91,Evrard+96}.  However, this cannot
extend from clusters down to groups without vastly overproducing the
extragalactic soft X-ray background.  Indeed, groups are observed to have a
smaller X-ray luminosity compared to the self-similar expectation yielding a
steeper scaling with temperature, $L_x\propto T^3$ \citep{Markevitch98}.  This
can be explained if the gas was initially preheated to an entropy floor of
$\sim100~\keV\,\cm^2$ which reconciles the simulated background with
observations \citep[see][for a review]{Voit2005}.

In principle, there are three physical processes that could produce such an
entropy floor for groups, all of which may occur in practice, raising the
question of which, if any, is dominant: (1) catastrophic cooling and collapse of
the low-entropy gas at the centers of halos, allowing accretion-shock heated gas
to adiabatically flow inward and replace the condensed gas with its elevated
entropy level \citep{Voit+2001,Voit+2001Nature,Voit+03}, (2) an early epoch of
global entropy injection prior to the formation of groups and clusters,
typically referred to as ``preheating''
\citep[e.g.,][]{Kaiser91,Evrard+91,Ponman+1999,Balogh+1999,Pen1999,Borgani+2001,Bryan+2001,Croft+2001,Bialek+01,Babul+2002,Voit+03,Voit+2005,BorganiViel2009,Stanek+2010},
and (3) self-regulated AGN feedback at the cores of groups and clusters
\citep[e.g.,][]{Churazov2001,Sijacki+2006,Sijacki+2007,Sijacki+2008,McNamara+07,Puchwein+2008,Booth+2009,McCarthy+2010,Dubois+2010,Teyssier+2011}.

Catastrophic cooling has been found to be unstable in large-scale numerical
hydrodynamic simulations, typically resulting in an untenably large fraction of
baryons in stars \citep[see][for a review]{Borgani+2009}.  More importantly,
runaway cooling generally precludes the existence of the observed hot groups,
making it clear that it cannot be the sole explanation of the low X-ray
luminosity of groups. However, radiative cooling is a physical process that will
occur if it is not perfectly balanced by some heating process and the
observations of multiphase gas in cool cores (CCs) suggest that cooling is
happening \citep{Donahue+2011}, presumably through thermal instability
\citep{McCourt+2011}.  Hence, while {\em some} of the elevated entropies in
groups have to be attributed to cooling, the absence of a cooling catastrophe
implies a (non-gravitational) energy feedback process that will inevitably also
raise the core entropy.

In contrast, preheating has met with broad success in explaining the observed
$L_x-T$ correlation, ostensibly by suppressing the formation of dense cluster
cores at low masses and therefore reducing $L_x$ for group-sized
objects. Typically, the redshift range adopted for the injection of the entropy,
$z\gtrsim3$, ensures that preheating is complete well before any galaxy group or
cluster has turned around from the Hubble expansion and started to collapse.
Hence the entropy, $K_e = kT n_e^{-2/3}$ (where $n_e$ is the electron density
and $k$ is the Boltzmann constant), is generally injected at the lowest possible
gas density (and thus temperature) at that redshift.  This minimizes the heating
necessary to produce the observed impact upon the entropy, typically $1\,\keV$
per particle.  Numerical simulations incorporating a minimum entropy of the gas,
or entropy floor, of $K_e \simeq (100$--$200)\,\keV\,\cm^{2}$ at redshifts
around $z\simeq3$--$4$, are able to steepen the $L-T$ relation from the purely
gravitational collapse estimate ($L_x\propto T^2$), and find broad agreement
with observations
\citep{Kaiser91,Balogh+1999,Bialek+01,Voit+02,Voit+03,Stanek+2010}.
Nevertheless, there are serious observational difficulties confronting the
standard preheating scenarios, apart from the theoretical problem of lacking a
physical model for the heating process.  First, after preheating, star formation
in $L_*$ and lower mass galaxies is significantly suppressed
\citep[e.g.,][]{Oh+2003,Benson+2003} and hence inconsistent with observational
data that finds, e.g., the peak of the star formation rate at redshifts $z\sim
2$.  Second, the presence of a high entropy floor everywhere in the universe
makes it impossible for groups to radiate away the excess entropy and impossible
to explain the existence of a subset of X-ray luminous, CC groups with steep
entropy profiles, and low values of the central entropy
\citep{McCarthy+08,Fang+08,Cavagnolo+2009}.

Heating via feedback processes, e.g., from star formation (stellar winds and
supernova) and AGNs (for a review, see \citealt{McNamara+07}), has already been
shown to have a large impact on the formation and history of galaxy clusters and
groups.  Self-regulated, inhomogeneous energy feedback mechanisms are very
successful in globally stabilizing the group and cluster atmospheres, and in
particular, preventing the cooling catastrophe.  The resulting gas mass
profiles, gas fractions, and the $L_x-T$ correlation compare impressively to
those observed \citep{Sijacki+2008,Puchwein+2008}.  While there seems to be a
globally convergent scenario emerging with convincing energetics and duty
cycles, the actual heating process has yet to be identified and may involve
interesting astrophysics, e.g., turbulence \citep{Ensslin+2006}, cosmic rays
\citep{Guo+2008,Ensslin+2011}, or plasma instabilities \citep{Kunz+2011}.

As we will show in this paper, the entropy injection by blazars is in some sense
an amalgam of both the preheating and feedback mechanisms.  Unlike typical AGN
feedback, the effect of the blazar heating is not localized, operating at much
larger distances typically and thus over considerably longer timescales.  Unlike
instantaneous preheating, the blazars provide a {\em time-dependent} entropy
injection rate, peaking near $z\sim1$.  As a consequence, we will show that the
formation of $L_*$ galaxies is not suppressed and early-forming groups are
little affected by blazar heating, having had time to cool and develop an X-ray
luminous {\em CC}; avoiding the primary difficulties with the
standard preheating scenario. On the other hand, the blazar-heated ICM ending up
in late-forming groups will not have had sufficient time to radiate the
additional entropy away before it can be gravitationally reprocessed in merging
shocks which are ubiquitous in a hierarchically growing universe. We will argue
that this can lead to elevated entropy core values resembling those of {\em
  non-cool core (NCC) clusters}.

\subsection{The dwarf problem in our Galaxy and nearby voids}
\label{sec:intro_dwarf}

The $\Lambda$ cold dark matter ($\Lambda$CDM) concordance cosmology predicts
that Milky Way-sized halos should contain many more dwarf-sized halos than the
observed number of dwarf galaxies, the so-called ``substructure problem'', or
within the local context the ``missing satellites problem'' \citep[for a recent
review, see][]{Krav:10}. Closely related to this problem is the ``void
phenomenon'' which is the apparent discrepancy between the number of dwarfs in
low-density regions in simulations and the paucity of those in observations
\citep{Peebles2001}.  In principle, both problems can be solved in three general
ways: by suppressing the formation of dwarf halos, suppressing the formation of
galaxies within existing dwarf halos, and/or suppressing star formation within
dwarf galaxies.  In the case of the ``void phenomenon'', there is a forth class
of models that solves the problem. Considering interacting dark matter and dark
energy as mediated, e.g., by a Yukawa coupling, implies a fifth force that
reduces late accretion onto halos and pushes matter out of voids, hence
resolving the discrepancy of dwarf abundances in voids \citep{Farrar+2004,
  Nusser+2005}.  In the following, we review the three classes of models that
are able to solve both problems simultaneously.

Suppressing the formation of dwarf halos is difficult to accomplish within the
context of $\Lambda$CDM, requiring modifications to the standard cosmological
paradigm, such as interacting dark matter \citep{Sper-Stei:00}, modifications to
the seed perturbation spectrum \citep{Zent-Bull:03}, or warm dark matter
\citep[WDM][]{Dalc-Hoga:01,Maccio_WDM2010}.  While we will discuss the last of
these in Section \ref{sec:WDM}, given the current success the $\Lambda$CDM model
has had predicting the observed halo structures \citep[see,
e.g.,][]{Dala-Koch:02,Mao_etal:04}, we will not comment upon these possibilities
further.

Suppressing the formation of dwarf galaxies, i.e., preventing the accretion of
baryons by existing dwarf halos, may be accomplished in principle by a variety
of mechanisms, including photoionization heating
\citep{Efst:92,Kauffmann+1993,Quin-Katz-Efst:96,Thou-Wein:96,Kity-Ikeu:00,Bullock+2000,Bullock+2001,Chiu+2001,Somerville2002,Dijk-Haim-Rees-Wein:04}
or accretion shock heating
\citep{Scan-Thac-Davi:01,Kravtsov+2004,Sigw-Ferr-Scan:05}.  If baryons can
collect, they may not be able to cool efficiently due to a lack of H {\sc i} as a result
of photoionization \citep{Haim-Rees-Loeb:97,Haim-Abel-Rees:00} or intrinsically
low metallicities
\citep{Kravtsov+2004,Kauf-Whee-Bull:07,Tass-Krav-Gned:08,Robe-Krav:08,Gned-Tass-Krav:09}.
Nevertheless, recent numerical simulations which self-consistently include the
photoionizing background due to star formation have found that while it does
have a pronounced effect, it cannot suppress dwarf galaxy formation at the level
implied by observations \citep{Hoeft+2006,Okamoto+2008,Nick_etal:11}.  More
importantly, the metallicity distribution of some dwarfs is inconsistent with
dwarfs generally being pre-reionization relics
\citep{Dolphin+2005,Fenn_etal:06,Holtzman+2006,Oban+2008}, and therefore the
suppression of dwarf formation must have occurred more recently than the epoch
of reionization.

Once formed, the gas may be removed from dwarf galaxies via photo-evaporation
\citep{Bark-Loeb:99,Shap-Ilie-Raga:04} and feedback from the first supernovae
\citep{MacL-Ferr:99,Deke-Woo:03,Mash-Wads-Couc:08,Jubelgas+2008,Wadepuhl+2011,Nick_etal:11,Uhlig+2012}.
Tidal interactions of satellite dwarf halos with the Milky Way may result in a
dramatic decrease in their mass, and to a lesser extent in circular velocity,
after $z\sim2$ \citep{Kravtsov+2004,Nick_etal:11}.  However, these processes
also strip material from larger halos, with the result that the smallest dwarf
spheroidal galaxies presently within the Local Group may have had a mass at
formation that was much larger than currently observed, and thus were capable
of building up a sizeable stellar component in their seemingly shallow potential
wells.

As we will show in this work, the heating due to blazars provides an additional
mechanism to suppress the formation of dwarfs.  Unlike photoionization models,
which typically invoke the heating at reionization, blazar heating provides a
well defined, time-dependent suppression mechanism, with the suppression rising
dramatically after $z\sim2$.  In addition, due to its insensitivity to density,
the heating from blazars suppresses structure formation most efficiently in the
low-density regions that are responsible for late-forming dwarf halos.  As a
result, the impact from blazars is not degenerate with variations in the
parameters of reionization and/or tidal interactions.

\subsection{Structure of this Paper}

This is the third in a series of three papers that discuss the potential
cosmological impact of TeV emission from blazars.  Paper I provides a
plausible mechanism for the local dissipation of the TeV luminosity, effectively
producing an additional heating process within the IGM, and its implications
upon high-energy gamma-ray observations.  Paper II estimates the magnitude of
the new heating term, describes the associated modifications to the thermal
history of the IGM, and shows how this can explain some recent observations of
the \Lya forest.  Paper III, this paper, considers the impact the new heating
term will have upon the structure and statistics of galaxy clusters and groups,
and upon the ages and properties of dwarf galaxies throughout the universe,
generally finding that blazar heating can help explain outstanding questions in
both cases.

In Section~\ref{sec:entropy}, we show that blazar heating necessarily implies
the injection of a tremendous amount of entropy and employ the implication for
the evolution of the warm and hot phases of the IGM.  In particular, we show the
broad implications of this heating on the X-ray population of clusters
(Section~\ref{sec:X-ray}), the entropy structure of galaxy groups/clusters
(Section~\ref{sec:groups}), the CC/NCC cluster bimodality
(Section~\ref{sec:bimod}), and its impact on the Sunyaev-Zel'dovich (SZ)
power-spectrum (Section~\ref{sec:SZ}).  We discuss the impact of this heating on
the formation of structure in the universe in Section~\ref{sec:structure
  formation}, showing from the viewpoint of linear theory that the formation of
dwarf galaxies that end up in Milky Way-sized halos (Section~\ref{sec:MWMS}) as
well as those in voids (Section~\ref{sec:VD}) are suppressed at late redshift
($z \lesssim 2$).  Finally, we conclude in Section~\ref{sec:conclusions}.

\begin{figure*}
  \begin{center}
    \includegraphics[width=0.495\textwidth]{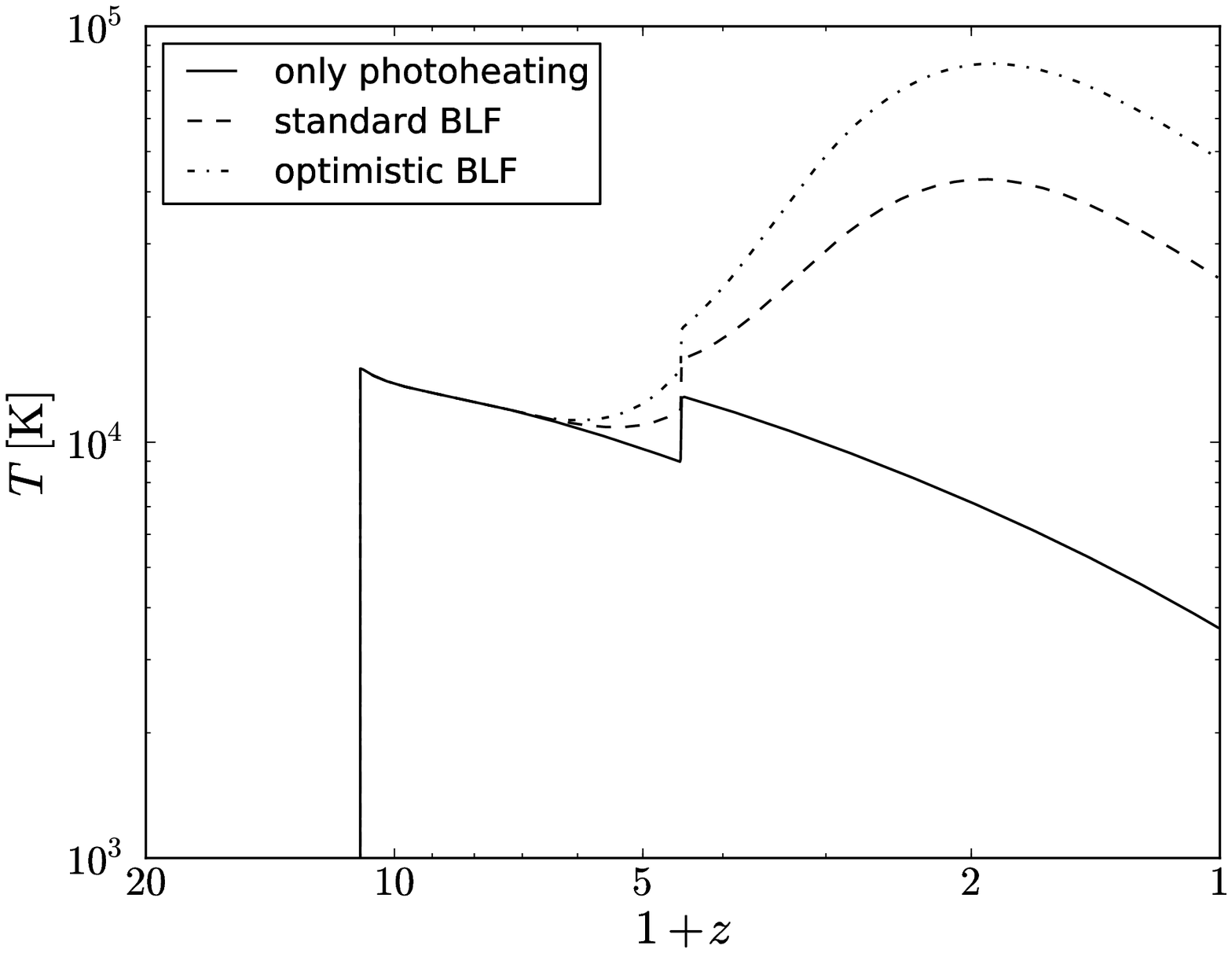}
    \includegraphics[width=0.495\textwidth]{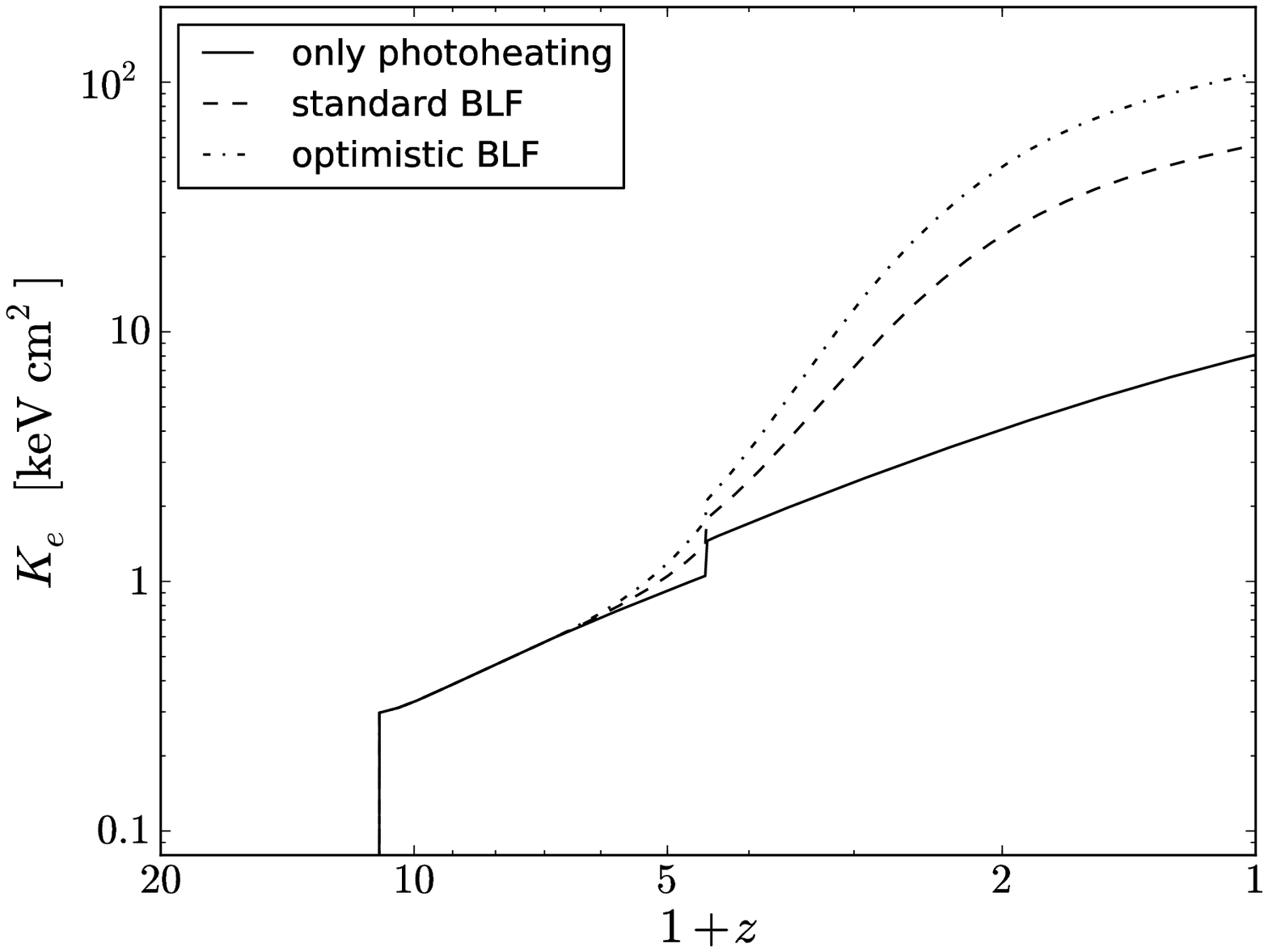}
  \end{center}
  \caption{Thermal history (left) and entropy history (right) of a patch of mean
    density ($\delta = 0$) of the IGM.  The solid curves are for pure
    photoheating with sudden reionization histories for H and He {\sc ii} at
    $z_{\rm reion} = 10$ and $z_{\rmn{He}\,\textsc{ii}} = 3.5$.  The effect of
    the ``loss of memory'' in the thermal history of the IGM is apparent in the
    entropy as the pure photoheating curve asymptotes to
    $K_e\approx8\,\keV\,\cm^2$ at $z=0$. The dashed (dash-dotted) lines show
    the evolution for the standard (optimistic) blazar heating model that
    employs the blazar luminosity density, i.e., using the redshift evolution of
    the quasar luminosity density from \citet{Hopkins+07} and are normalized to
    the local heating rate.  At $z=0$, those differ by a factor of two and the
    standard one has been estimated from the observed number of blazars,
    employing a conservative estimate for the incompleteness correction, while
    the optimistic one naturally matches the inverted temperature--density
    relation in the \Lya data \citep{Bolton+08,Viel+09,Puchwein+2011}.}
   \label{f:thermal history} 
\end{figure*}

The calculations presented below assume the WMAP7 cosmology \citep{WMAP7_2011}
with $h_0 = 0.704$, $\Omega_\rmn{DM} = 0.227$, $\Omega_{B} = 0.0456$,
$\Omega_{\Lambda} = 0.728 $, $\sigma_8=0.81$, $n_s=0.967$, and a matter transfer
function that accounts for the baryonic features \citep{Eisenstein+1998}.

\section{The Evolution of Entropy and the Hot  Phase of the Intergalactic Medium}\label{sec:entropy}

The injection of large amounts of energy into the IGM by blazars is accompanied
by a substantial increase in the IGM entropy.  Here, we discuss the impact of
this additional entropy upon the evolution of the gravitationally heated warm
and hot components of the IGM.  In doing so we neglect radiative cooling, an
approximation that is well justified in the low-density regions of the IGM
generally and one that we will justify for late-forming groups and clusters
below.  In this case the entropy of the IGM necessarily increases during
gravitational collapse (e.g., due to structure-formation shocks, feedback,
and/or photoheating).  As a consequence, the entropy injected by TeV blazars
places an elevated floor upon the entropy of late-forming structures within the
IGM.  If radiative cooling is permitted (e.g., at the centers of early-forming
clusters), the gas entropy may decrease,\footnote{According to the second law of
  thermodynamics, the entropy of an isolated system which is not in equilibrium
  has to increase over time and can only remain constant during a reversible
  process. In the case of radiative cooling, the gas shares its entropy with the
  cosmic radiation field---a process that effectively removes entropy from the
  gaseous phase. However, the total entropy of the system including gas and
  radiation increases in this process.} reducing the impact of the blazar
heating.  In this work, we will encounter two different definitions of
entropy\footnote{These definitions of entropy are related to the standard
  thermodynamic definition of entropy per particle by $s=k\ln K^{3/2} + s_0$,
  where $s_0$ is a constant that depends upon fundamental constants and a
  mixture of particle masses.}  which are proportional to each other, namely,
\begin{equation}
  \label{eq:Ke}
  K_e\equiv \frac{kT}{n_e^{2/3}}\quad\mbox{and}\quad
  K \equiv \frac{P}{\rho_g^{5/3}} = \frac{kT}{\mu m_p \rho_g^{2/3}}.
\end{equation}
$K_e$ is a quantity conveniently used in the X-ray literature as it can be
directly constructed from the observables temperature, $T$, and electron
density, $n_e$. The quantity $K$ is the constant of proportionality in the
equation of state $P(\rho_g)$ for an adiabatic monatomic gas with mass
density $\rho_g$ and pressure $P$.

The evolution of the temperature and the entropy are shown in Figure
\ref{f:thermal history} for a fluid element at overdensity $\delta
=\rho/\bar{\rho}-1=0$ (where $\bar{\rho}$ and $\rho$ denote the average and
local matter density) for the cases of pure photoheating and two realizations of
blazar heating. Our standard model is normalized to the observed number of TeV
blazars, accounts for an incompleteness correction for incomplete sky coverage
at TeV energies and duty cycle, employs conservative assumptions of spectral
corrections and contributing source classes for blazar heating, and has a
heating rate of $7\times 10^{-8}\,\mbox{eV cm}^{-3}\,\mbox{Gyr}^{-1}$ at $z=0$
(Paper II). The heating rate in the optimistic model is a factor two larger,
implying a rate of $1.4\times 10^{-7}\,\mbox{eV cm}^{-3}\,\mbox{Gyr}^{-1}$ at
$z=0$ and matches the inverted temperature-density relation found in high-redshift \Lya
observations \citep{Bolton+08,Viel+09,Puchwein+2011}.  Since any mechanism that injects energy
will produce a corresponding increase in $K_e$, in the absence of cooling the
entropy accumulates monotonically regardless of the mechanism responsible for
heating the IGM.  Nevertheless, some general statements can be made about the
distinction between photoheating and that due to blazars.

Due to the ionization balance maintained between recombination and
photoionization, photoheating produces a generic entropy injection profile
following the epoch of reionization; this is simply the ``loss of memory''
effect \citep{Hui97,Hui03}.  In the absence of the blazar component,
photoheating of hydrogen is the main source of heating for the universe at mean
density at late times.  The heating rate is given by the rate at which hydrogen
recombines and is then reionized.  In photoionization equilibrium, the
temperature of the system will adjust such that the timescales of the net
photoheating and recombination cooling rates become equal implying a steady
state.  Since recombination does not depend on the UV photon density the rate of
photoionization is also independent of the UV photon density.  Hence the late
time photoheating of the IGM is independent of the number of ionizing photons,
and thus the number of ionizing sources (above the ionization threshold).  As a
consequence, the photoheating contribution to the IGM entropy at $z=0$ is
limited to near $K_e\approx 8\,\keV\,\cm^2$ for $\delta=0$.

Such an argument does not work for blazar heating.  Here the efficiency is close
to the maximum of 100\% all the time\footnote{In general, the efficiency for
  heating depends on the saturated, nonlinear damping rate of the maximally
  growing mode of the dissipating plasma instability, $\Gamma_\rmn{M}$, and the
  inverse Compton cooling rate, $\Gamma_\rmn{IC}$, and scales as
  $\Gamma_\rmn{M}/(\Gamma_\rmn{M} + \Gamma_\rmn{IC})$. We assume that the
  nonlinear damping rate is equal (or of order) the linear growth rate (see
  Section 3.5 of Paper I). For the linear growth rate of the oblique
  instability, this efficiency is close to unity for luminous blazars (Paper
  I). Numerical simulations for a mildly relativistic pair beam penetrating into
  a hot, dense background plasma suggest that a significant fraction ($\sim
  20\%$) of the beam energy is heating the background plasma through the oblique
  instability before the two-stream plasma instability takes over, potentially
  dissipating another fraction of the beam kinetic energy
  \citep{Bret-Grem-Diec:10}.}  provided there are sufficient numbers of EBL
photons and each point in space is reached by a number of blazar beams of TeV
photons (which is likely the case; see Section 3.2 of Paper II).  Therefore, the
heating rate, and thus temperature of the IGM, is nearly linearly dependent upon
the VHEGR luminosity density of TeV blazars.  Correspondingly, the entropy
injection from blazars depends sensitively upon the history of the TeV blazar
population, and thus the cumulative contribution of blazars to the present-day
$K_e$ is somewhat uncertain.  Nevertheless, given our conservative estimate of
the blazar luminosity density in Paper I, fixing it to the quasar luminosity
density, we find that blazar heating raises the entropy substantially, starting
around the epoch of He {\sc ii} reionization ($z\sim3.5$) and by $z=0$, the inclusion
of blazar heating raises the entropy of the $\delta=0$ fluid element to
$K_e\approx(50-100)\,\keV\,\cm^2$, approximately an order of magnitude larger than the
case of photoheating alone.

\begin{figure}
\begin{center}
\includegraphics[width=\columnwidth]{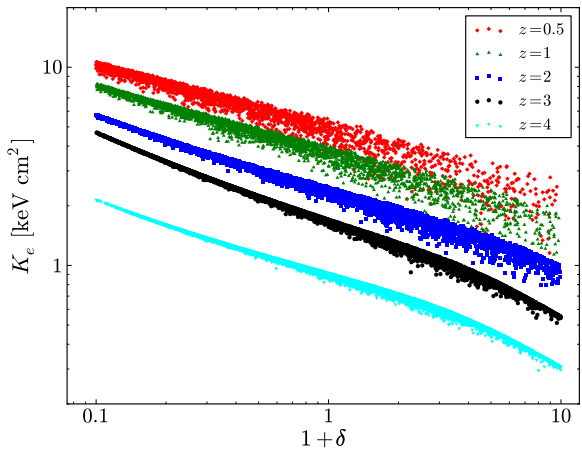}
\end{center}
  \caption{Entropy-density relation ($K_e$ versus $1+\delta$) for $z=0.5$
    (red diamonds), 1 (green triangles), 2 (blue squares), 3 (black
    circles), and 4 (cyan inverted triangles) for a pure photoheating model.}
   \label{fig:entropy eos}
\end{figure}

\begin{figure*}
\begin{center}
\includegraphics[width=0.495\textwidth]{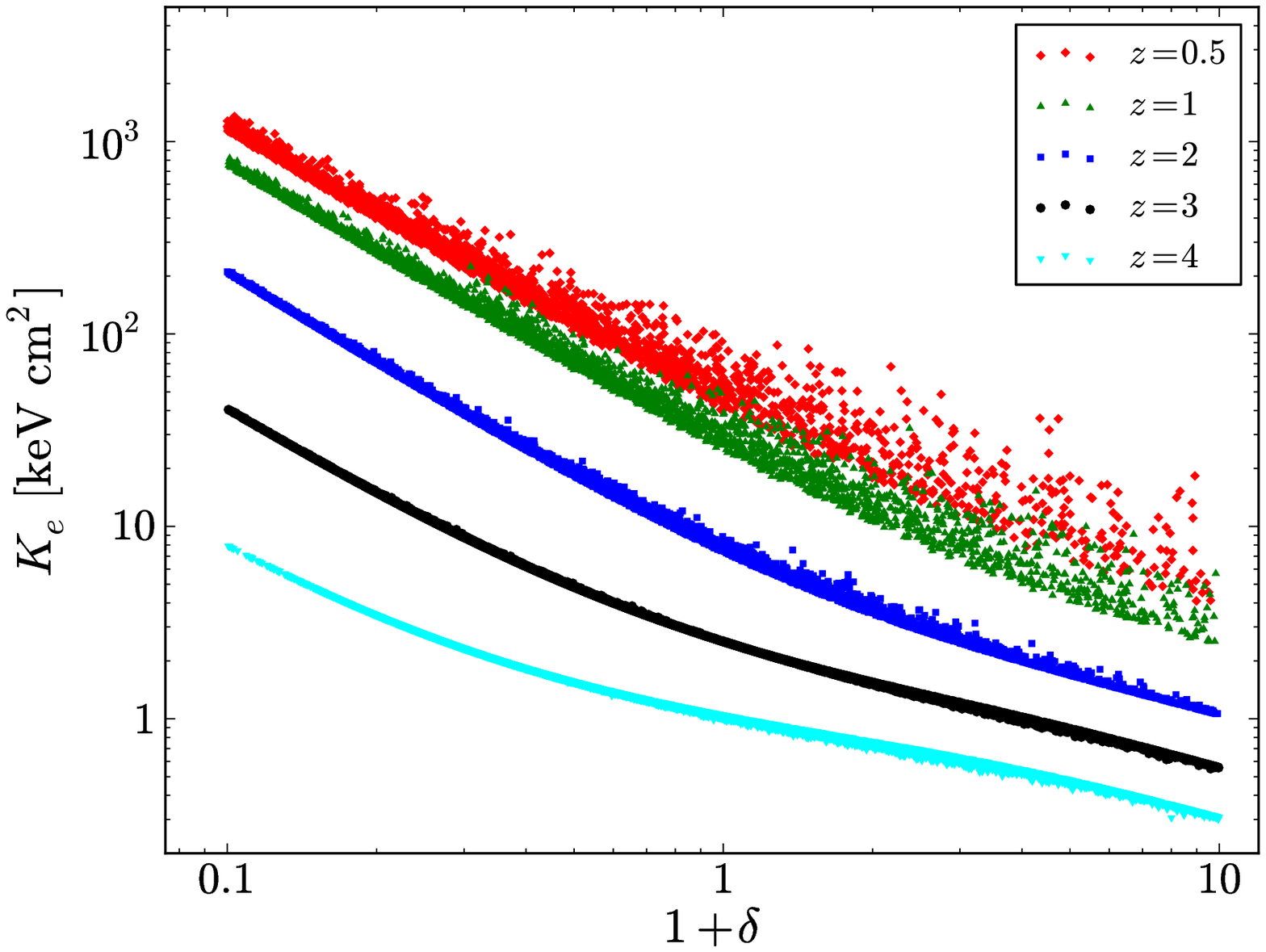}
\includegraphics[width=0.495\textwidth]{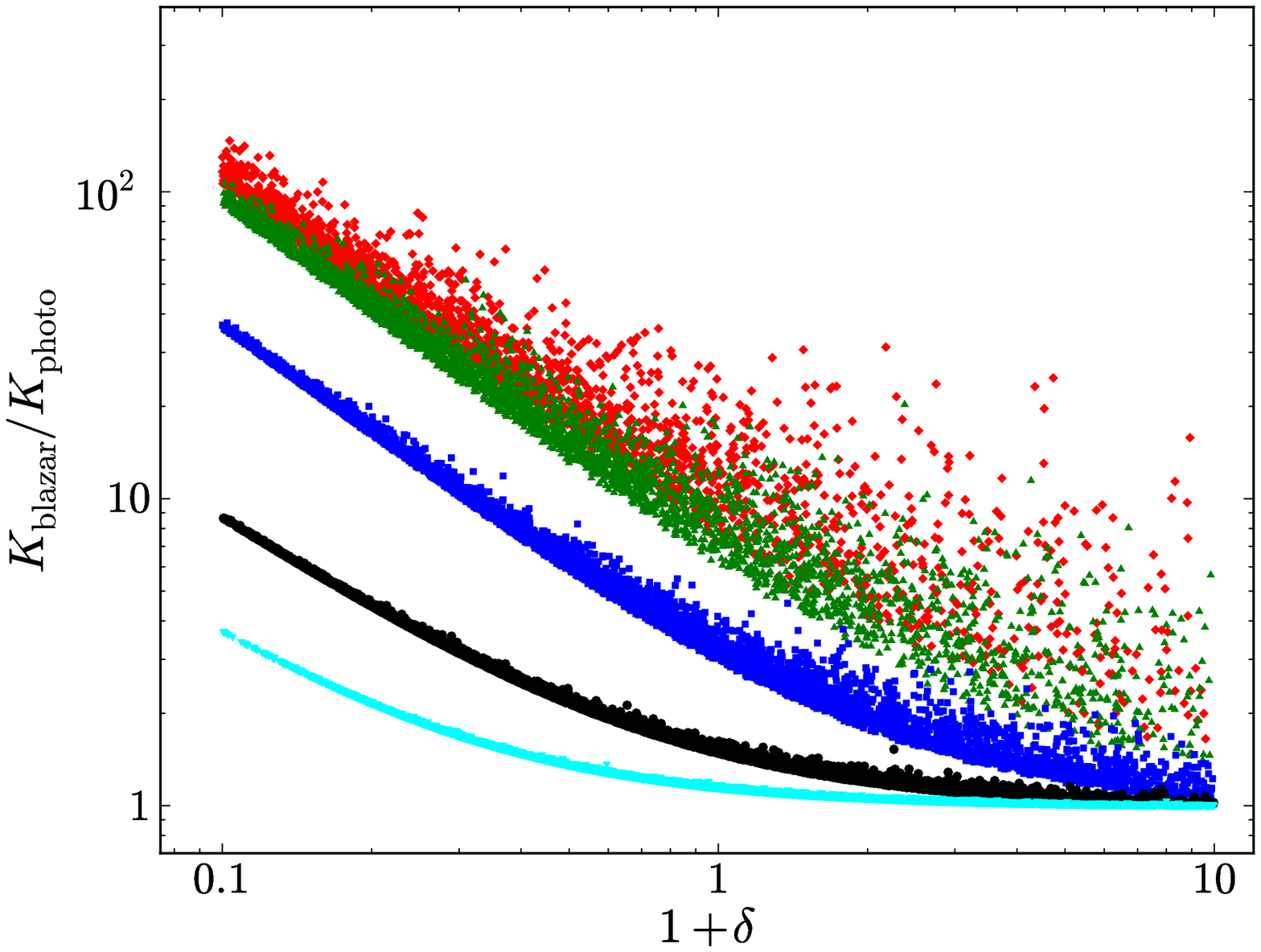}\\
\includegraphics[width=0.495\textwidth]{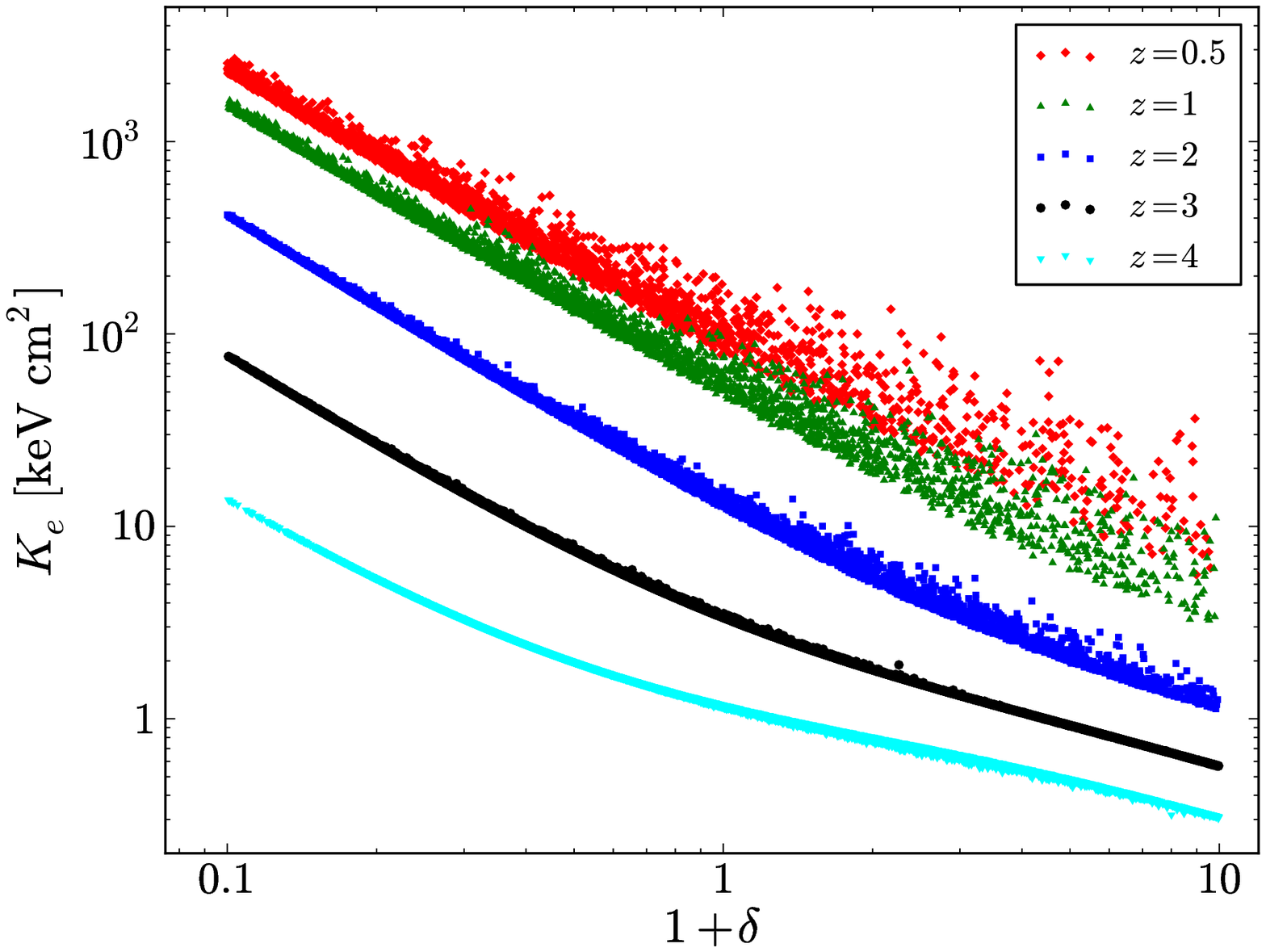}
\includegraphics[width=0.495\textwidth]{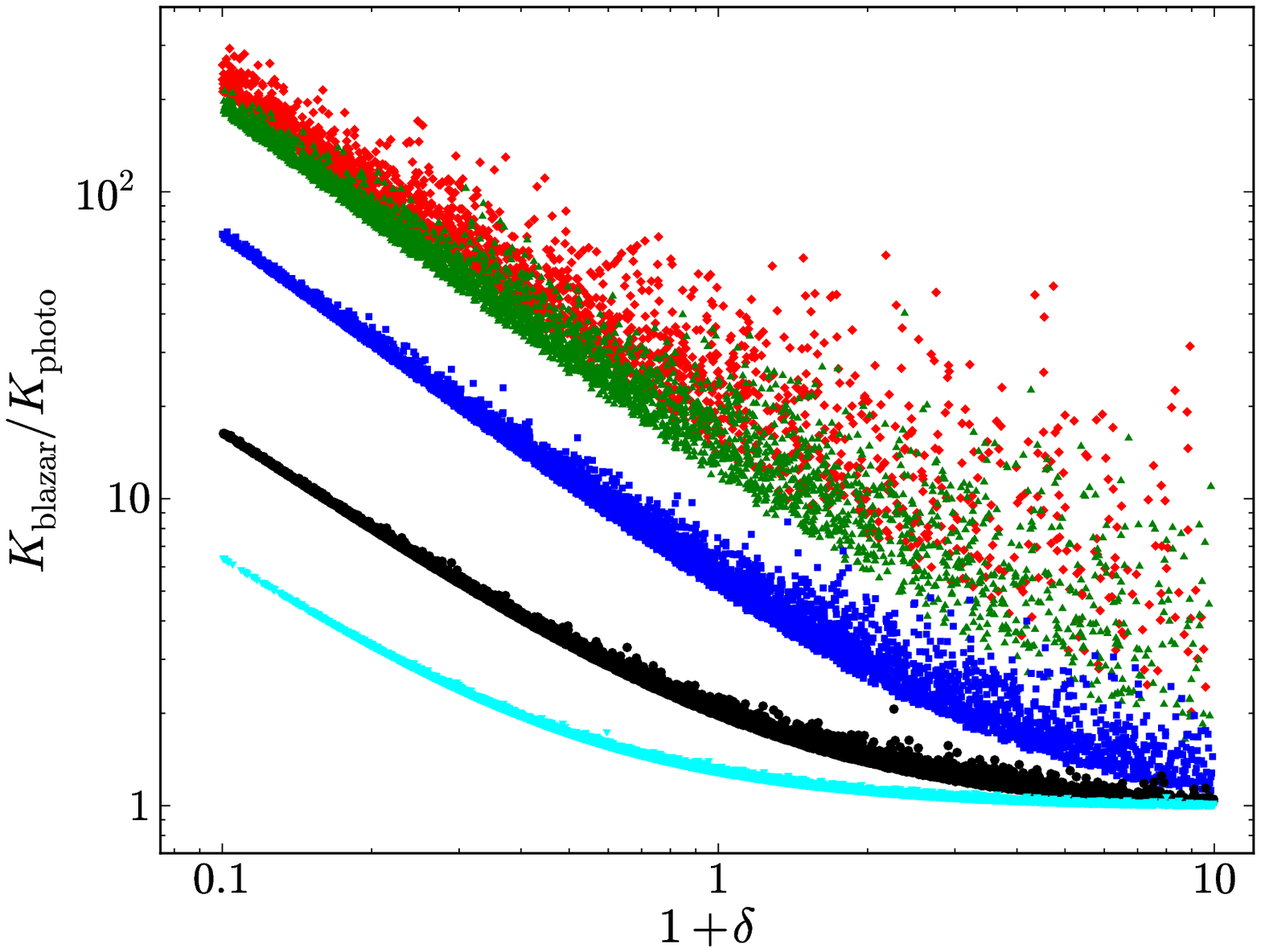}\\
\end{center}
\caption{Entropy-density relation ($K_e$ versus $1+\delta$, left) for the
  standard (top) and optimistic (bottom) blazar heating model. Blazar heating
  increases the entropy over the pure photoheating case by a factor
  $K_\rmn{blazar}/K_\rmn{photo}$ which is shown versus $1+\delta$ (right) again for
  the standard (top) and optimistic (bottom) blazar model. Shown are different
  redshifts $z=0.5$ (red diamonds), 1 (green triangles), 2 (blue squares), 3
  (black circles), and 4 (cyan inverted triangles).}
   \label{fig:entropy blazar}
\end{figure*}

As discussed in Paper II, the magnitude of the temperature enhancement due to
blazar heating is density dependent: while the photoheating rate depends
linearly on density, blazar heating is independent of density and dominates
photoheating for regions with $\delta \lesssim 10$.  Thus, blazar heating is
expected to have the largest effects in voids.  In Figures \ref{fig:entropy eos}
and \ref{fig:entropy blazar} we show the entropy-density relation ($K_e$
vs.{\ }$1+\delta$) for a variety of redshifts, ranging from $z=0.5$ to $z=4$.  For
comparison, Figure \ref{fig:entropy eos} shows the case of photoheating only:
the entropy-density relation reflects the effect of photoionization which
deposits a uniform energy per baryon during H and He {\sc ii} reionization implying
$K\propto(1+\delta)^{-2/3}$.  This effect is quickly erased due to the ``loss of
memory'' effect as well as adiabatic cooling due to the Hubble expansion
approaching an asymptotically constant entropy-density relation at late times.

Figure \ref{fig:entropy blazar} shows the effect of TeV blazar heating on the
entropy-density relation for our standard (top) and optimistic (bottom) model.
As expected, blazars have the most dramatic effects upon $K_e$ in low-density
regions ($1+\delta\simeq0.1$) due to the spatially uniform and IGM density
independence of their volumetric heating rate.  This {\em steepens} the
entropy-density relation, causing it to approach an asymptotics of $K\propto
(1+\delta)^{-5/3}$ in voids.  In fact, blazar-induced entropies can exceed those
due to photoionization by a factor of $125$--$250$ in these low-density regions
($1+\delta\simeq0.1$, see right-hand panels of Figure \ref{fig:entropy blazar}),
reaching $K_e\simeq(1250$--$2500)\,\keV\,\cm^2$ by $z=0.5$.  For high density
regions ($1+\delta\simeq10$) the effect is still pronounced, increasing the IGM
entropy by up to an order of magnitude, i.e.,
$K_e\approx(10$--$20)\,\keV\,\cm^2$ at $z=0.5$.

Apart from increasing the mean entropy, blazar heating also increases the
scatter in entropy at fixed density which is especially noticeable for $z=0.5$.
Fractionally, this scatter appears to be roughly twice as large when blazar
heating is included, though in absolute terms the scatter is much larger for the
blazar cases.  This is because even in linear theory, the entropy (and
temperature) of any patch of overdensity $\delta$ depends on its ``collapse''
history.  Namely, whether it collapses like a Zel'dovich pancake, i.e., in one
dimension first which implies a smaller entropy and temperature, or more
spherically which produces a greater entropy and temperature, i.e., more akin to
accretion onto a local overdensity.

The fractional scatter in $K_\rmn{blazar}/K_\rmn{photo}$ is much larger than
that found in either entropy individually.  The reason is that the scatter
induced by blazar heating is anti-correlated with that induced by photoheating
due to their very different dependencies upon IGM density.  Patches that begin
at low densities when blazar heating is ignored are biased toward lower
entropies as a consequence of an extended period of slow recombination.  When
blazar heating is included, these same patches are more efficiently heated, and
therefore biased toward higher entropies.

The redshift evolution of blazar heating introduces additional stochasticity:
patches at a given density sample a distribution of turnaround times and hence
preheated entropy values. After turnaround, the gas is adiabatically
compressed and moves on an adiabat (a line of constant $K_e$) to higher density
with an entropy value that depends on the ``collapse'' time, and the steeper
entropy-density relation generates the larger scatter at any $\delta\gtrsim1$.
Hence by preheating the universe, the first shells of gas that are able to
collapse onto an object do not experience shock heating as they are only
adiabatically compressed which results in an entropy floor of the object after
formation. The later collapsing shells experience weaker shocks with a smaller
Mach number due to their already elevated entropy level.

Associated with the large increase in the entropy of the IGM are a number of
observational effects.  In the following we discuss the impact blazar heating
has upon the correlation between cluster X-ray luminosity and temperature, the
entropy profiles and the CC/NCC bimodality of clusters and groups, and the SZ
power spectrum.  In each we note the unique effects imposed by the relatively
recent nature of the entropy injection.

\subsection{Implications for the X-ray emission of groups and clusters}
\label{sec:X-ray}

Blazar heating raises the entropy floor for $z\lesssim2$ with its peak
contribution around $z\sim1$ and acts on scales much larger than the turnaround
region of clusters. The formation epoch of groups and clusters is roughly
matched to the epoch of heating.  As a consequence, early-forming, i.e., old
groups may be little affected by blazar heating, had time to cool and develop an
X-ray luminous core, potentially representing the class of CC
groups/clusters. On the other hand, the blazar-heated IGM collapsing into
late-forming, young groups will not have had sufficient time to cool. These
groups will ``remember'' the elevated entropy floor as an extended core that
substantially changes the initial conditions for their subsequent hierarchical
evolution.  However, because the heating occurs at late times, and thus after
the first groups and clusters have begun to form, a larger energy input is
required in comparison to the traditionally employed early preheating models.

For the observability of such a non-gravitational entropy floor and its
implication on the thermal history of the ICM, it is important to consider its
cooling timescale for typical entropy values of the IGM around $z\simeq 0.5$ and
parameters typical of the intra-group medium:
\begin{multline}
  t_\mathrm{cool} =  \frac{3 n kT}{2 n_e n_\rmn{H} \Lambda(T,Z)} =
  4.5\,\Gyr
  \times \left(\frac{K_e}{75\,\keV\,\cm^2}\right)^{3/2 }\\
  \times \left(\frac{kT}{1\,\keV}\right)^{-1/2} 
  \left(\frac{\Lambda(kT,Z)}{\Lambda(1\,\keV,0.3Z_\sun)}\right)^{-1}.
  \label{eq:tcool2}
\end{multline}
Here $n=2.2 n_\rmn{H}$, $\Lambda(T,Z)$ is the cooling function at a given $T$
and metallicity, $Z$, \citep[which is a relatively flat function of $T$ near
$kT=1\,\keV$ and $0.3Z_\sun$,][]{Sutherland+1993}.  A look-back time of
$4.5\,\Gyr$ corresponds to a redshift of $z=0.425$.  

Instantaneous preheating by blazars (see Figure~\ref{f:thermal history}) is
incapable of {\em directly} producing the highest entropies (up to
$600\,\keV\,\cm^2$) that are presently observed in a few clusters
\citep{Cavagnolo+2009}.  Comparing the cooling time to the dynamical time of a
cluster/group of approximately $\sim1$~Gyr or the typical timescales of
significant turbulent pressure support of $\sim2$~Gyr after a cluster merger
which provides continuous heating throughout \citep{Paul+2011}, we expect our
heating to {\em indirectly} impact the entropy distribution through
gravitational reprocessing of the blazar preheated core entropy, leading to a
gravitational amplification of entropy that we now explain.

A larger central entropy, or equivalently, a lower central density\footnote{For
  a cluster with a given mass, a larger core entropy implies a lower gas density
  as the (constant) core temperature $kT\propto K n^{2/3}$ reflects the virial
  value imparted by the cluster's potential depth.}, of a merging cluster
facilitates shock heating which implies an increase of the core entropy of the
final object compared to that of a dense cooled core.  CCs have a large
inertia causing them to be rather resilient against shock heating, typically
surviving the merger with only a marginally elevated entropy level which is then
subject to fast radiative cooling.  This effect gives rise to the well-known
overcooling problem in cosmological simulations of galaxy clusters if cooling is
not counteracted by any feedback process \citep[e.g.,][]{Borgani+2009}.

In order to quantify the effect of gravitational reprocessing of a preheated
entropy core, let us define a net entropy amplification factor through shock
heating. This is easiest done by comparing two types of simulations of galaxy
cluster formation: one with gravitational physics only and one where
gravitational physics is supplemented by a preheating epoch of the IGM at mean
density with a uniform entropy floor of $K_\rmn{floor}$ which precedes the
turnaround and formation of a galaxy cluster.  In the first case, the core
entropy structure is set by gravitational formation shocks leading to a central
entropy of $K_{\rmn{grav},0}$. In the preheating case, the core entropy is
raised to an elevated level after nonlinear evolution during virialization,
$K_{\rmn{pre},0}$. Hence, the net entropy amplification factor is then defined
as the ratio of the core entropy in the preheating case, $K_{\rmn{pre},0}$, to
that obtained by adding the IGM entropy floor value to the entropy in
gravitational heating case only, $K_\rmn{floor}+K_{\rmn{grav},0}$.

Non-radiative cosmological simulations of galaxy clusters and groups demonstrate
that the net entropy amplification factor can reach values ranging from 3 to 5
for clusters and groups, respectively, in the high-entropy case with a
preheating value of $K_\rmn{floor}=100\,\keV\,\cm^2$ imposed at $z=3$ \citep[see
Figures 1 and 2 in][]{Borgani+2005}.  For the low-entropy case with
$K_\rmn{floor}=25\,\keV\,\cm^2$, this amplification factor is only mildly
reduced to $2-5$ for clusters and groups, respectively.  This entropy
amplification factor seems to be reduced for radiative cluster simulations that
include cooling and star formation, possibly due to the assumed early epoch of
instantaneous entropy injection at $z=3$ that facilitates cooling of the entropy
floor thereafter \citep[see Figures 3 and 4 in][]{Borgani+2005}. Groups,
however, still remain severely affected by an early epoch of preheating.  Thus
in combination with gravitational amplification, blazar heating could have an
important effect on the subsequent thermodynamic evolution of late-forming
groups and possibly the clusters they evolve into. This, however, is subject to
nonlinear structure formation and needs to be carefully studied with
cosmological hydrodynamic simulations containing a large sample of
well-resolved galaxy clusters which sample the full distribution of formation
redshifts.

\begin{figure}
\begin{center}
\includegraphics[width=0.495\textwidth]{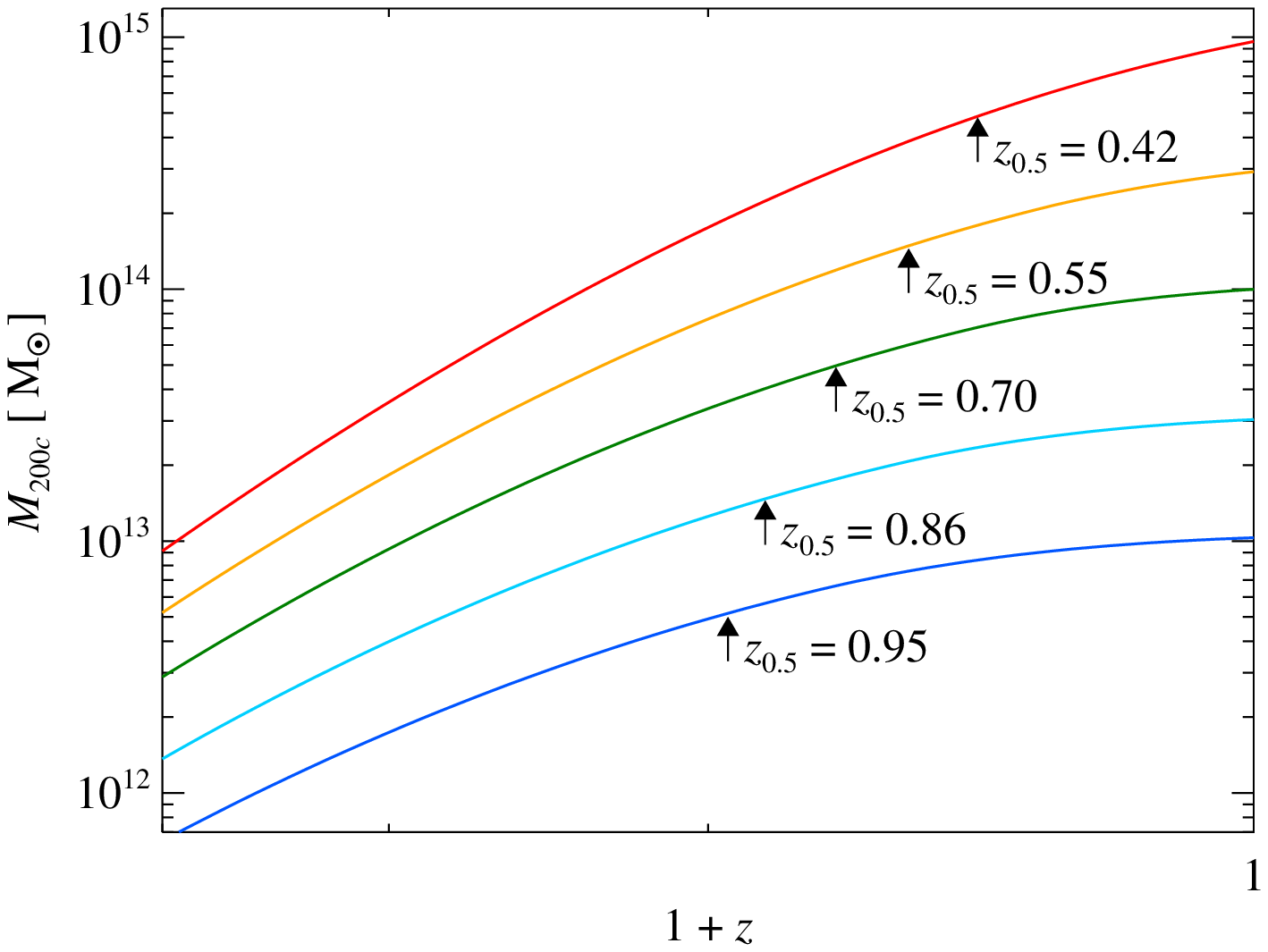}\\
\includegraphics[width=0.495\textwidth]{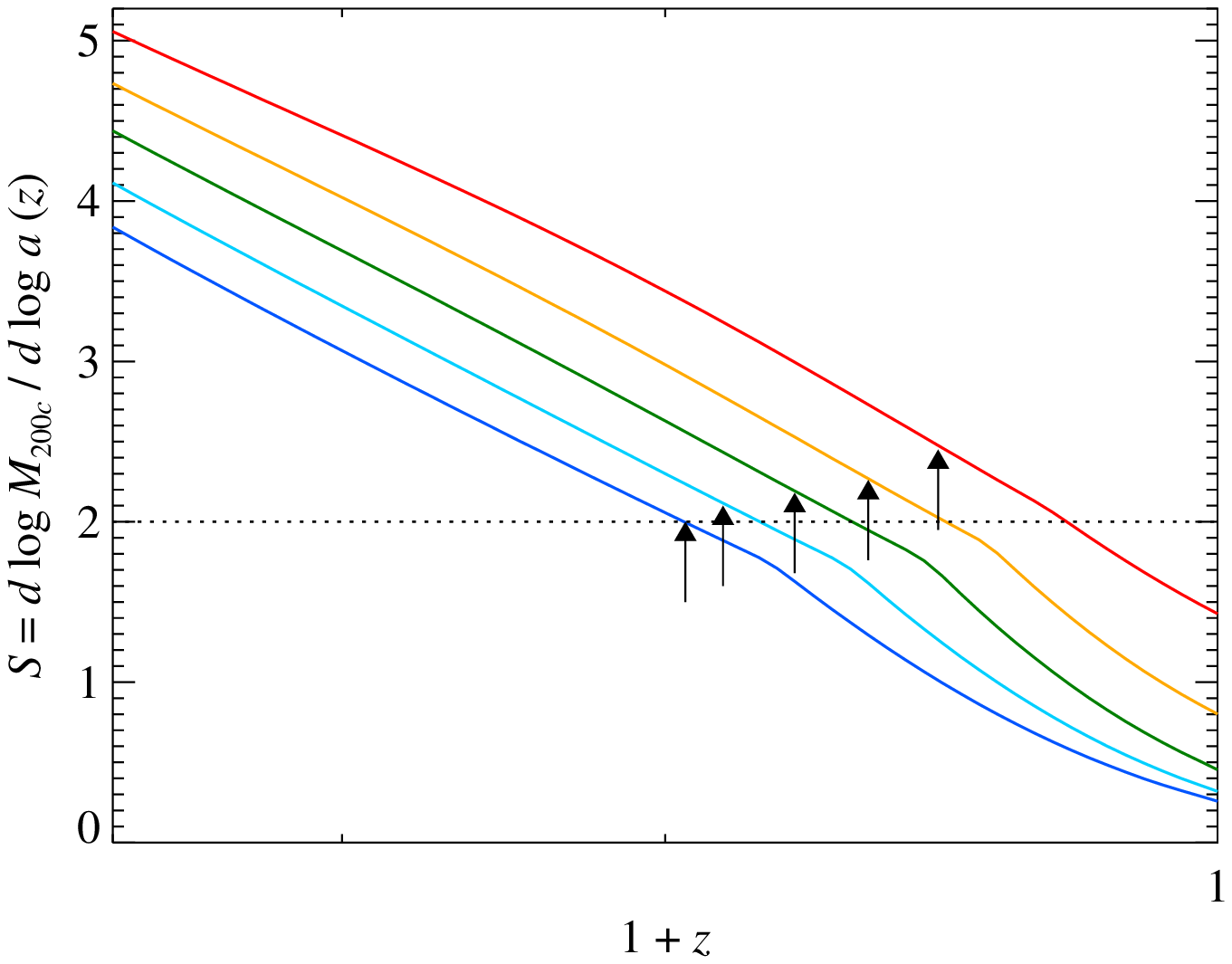}
\end{center}
\caption{Median mass accretion histories (top) and mass accretion rates defined
  by the logarithmic slope $S=d \log M_{200\,c} / d \log a(z)$ (bottom) for five
  characteristic halo masses $M_{200\,c}$ according to the model by
  \citet{Zhao+2009}. Arrows label the half-mass redshifts of the systems.  At
  late times, the mass accretion rates on group scales start to flatten out
  which is a result of the change from the fast to the slow accretion mode. In
  contrast, clusters are still forming today as the half-mass redshift $z_{0.5}$
  are still in the fast accretion regime characterized by $S>2$.}
   \label{f:macc}
\end{figure}

The impact of the entropy injection due to blazar heating upon clusters is
somewhat less clear.  Clusters form in highly biased regions through the mergers
of smaller, virialized systems such as groups.  In most cases, these need to
have already formed by $z=1$ which can be inferred from the mass accretion
history in Figure~\ref{f:macc}. This shows that on average, the most massive
group progenitor of a cluster with virial mass\footnote{In this Section, we
  define the virial mass $M_{200\,c}$ as the mass of a sphere enclosing a mean
  density that is 200 times the {\em critical} density of the universe.}
$M_{200\,c}=10^{14}\,\rmn{M}_\sun$ crossed a mass threshold of $3\times
10^{13}\,\rmn{M}_\sun$ before $z=1$. At this time the entropy floor due to
blazar heating was still smaller with typical values
$K(\delta=0,z=1)\approx(25$--$50)\,\keV\,\cm^2$ implying a smaller entropy core
in groups at higher redshifts $z\gtrsim1$. Hence, we naively may not expect
blazar heating to have a large impact upon the X-ray luminosity and entropy
profiles of clusters. However, this is not the case for two reasons.

First, the mass accretion rate, defined by the logarithmic slope $S=d \log
M_{200\,c} / d \log a(z)$, is larger for more massive systems and at higher
redshifts (where we have introduced the cosmic scale factor, $a\equiv1/(1+z)$;
see bottom panel of Figure~\ref{f:macc}). The mass accretion history of a
halo observed at $z=0$ grows on average as $M(z) = M_0\,\exp(-\xi z)$
\citep{Wechsler+2002}.  The single free parameter in the model, $\xi$, can be
related to a characteristic formation redshift $1+z_c = \xi/S$.  Hence we
find ourselves in the fast accretion regime if the mass accretion rate is larger
than a characteristic value usually taken to be $S=2$.  In particular, the
relative mass accretion rates increase from $z=0$ to 2 by a factor 3 for
clusters ($M_{200\,c}=10^{15} \rmn{M}_\sun$) and 10 for groups
($M_{200\,c}=10^{13} \rmn{M}_\sun$) \citep[see also][]{Gottloeber+2001}. A
larger relative mass accretion rate implies that the gas in the core is heated
by accretion processes at a faster rate.  This means that a lower preheated
core entropy value early-on can in principle be gravitationally processed at a
substantially higher rate and, as a result of this, move onto a higher adiabat
with a longer cooling time.

Second, blazars like AGNs follow the clustering bias of the matter density field
and hence turn on first in highly biased regions, i.e., regions that evolve into
clusters and super-clusters.  At late times ($z\lesssim3.5$) blazar heating is
expected to be nearly spatially uniform, since more than a single blazar
contribute significantly to the local heating rate of any given patch of the
universe (Paper II).  At much lower redshifts, the number of contributing
sources grows dramatically, approaching $10^3$ at $z=0$, and thus today blazar
heating is nearly homogeneous.  Prior to $z\simeq3.5$, however, blazar heating
may exhibit $\sim50\%$ fluctuations locally, due to the paucity of blazars in
the early universe. By $z\simeq6$ as much as 75\% of the local heating can be
due to a single object, implying large Poisson fluctuations in the heating
rate. This results in a clustering bias at early times.

The rare first blazars are expected to appear in highly biased regions,
corresponding to those that later evolve into groups and clusters, and therefore
themselves be clustered. The likelihood of a patch of the universe being heated
by such a blazar is then enhanced in the biased regions, where both the heating
rate is larger (since the VHEGR density is larger) and the probability of being
covered by a blazar in the cluster is larger.\footnote{In the limit of a few,
  highly clustered blazars, the VHEGR intensity declines exponentially from the
  clustering site for a spatially constant distribution of EBL photons, giving
  rise to an exponentially decreasing heating rate (on the scale of the VHEGR
  mean free path which is larger than the clustering length scale). However, if
  the distribution of EBL photons is also clustered and correlates spatially
  with the highly biased regions (which is expected in a hierarchically growing
  universe), the heating rate of those biased regions will be additionally
  enhanced in comparison to the void regions that have a low probability of
  being covered by a blazar. We stress that the effect of spatially and
  temporarily inhomogeneous blazar heating due to the clustering bias is
  expected to be absent at late times ($z\lesssim3.5$) due to the frequent
  occurrence of blazars.} It is this early preheating in groups that may
facilitate their evolution into NCC clusters (via the gravitational reprocessing
of the high-entropy cores).  Therefore, while we will focus upon
group/small-cluster mass scale in our study of the entropy structure immediately
after group/cluster formation, gravitational reprocessing combined with the bias
in the early blazars suggest that these effects may be important for more
massive clusters as well, propagating the effect of blazar heating up the mass
hierarchy.

The formation time of groups and clusters determines the blazar contribution to
their entropy, inducing an intrinsic scatter in the $L_x-T$ relation associated
with the distribution of collapse histories.  As a result, we would expect to
find systematically higher entropies in younger groups and possibly clusters
provided gravitational reprocessing was effective.  There is some evidence that
this is the case.  Optically selected, and therefore young, group and cluster
samples have on average lower X-ray surface brightness and smaller gas mass
fractions compared to X-ray-selected samples
\citep{Mahdavi+2000,Hicks+2008,Dai+2010}.  Optically bright, small to moderately
massive clusters ($kT>4\,\keV$) at redshifts $z=0.6$--$1.1$ are under-luminous
in X-rays for a given $T$, which implies a reduced gas density and by extension
an enhanced core entropy by roughly a factor of two \citep{Hicks+2008}.
Similarly, an X-ray stacking analysis of {\em ROSAT} data based on optically
selected groups at low redshift ($0.5\,\keV<kT<2\,\keV$) from the Two Micron All
Sky Survey catalog finds systematically lower gas mass fractions than expected,
$f_\rmn{gas}$, (within an over-density of 500 times the critical density of the
universe) and flatter temperature profiles \citep{Dai+2010}\footnote{We caution
  the reader that such a stacking analysis could have potential biases: the
  stacked X-ray spectrum in a given richness bin might be dominated by a few hot
  systems and fitting an average spectrum could then bias the temperatures and
  hydrostatic masses high and hence the gas fractions low. Eddington bias from
  more numerous smaller systems can additionally lower the average gas fraction
  at a given optical richness. Before drawing far reaching conclusions, careful
  mock analyses are needed to confirm these results.}, again implying larger
entropies.  Taken at face value, this is in conflict with careful X-ray studies
using {\em Chandra} data \citep{Sun+2009} {\em if} both samples are believed to
represent the same underlying distribution.  Alternatively, this is perfectly
consistent if the entropy of clusters and groups varies with formation time,
with the peak entropy injection occurring near $z\sim1$.  Both are well matched
to the properties of AGN feedback generally, and blazar heating especially.

\subsection{Thermodynamic structure of galaxy groups and clusters} \label{sec:groups}

The large injections of entropy necessarily influence the formation of cosmic
structure. While the incorporation of an instantaneous entropy floor in
numerical simulations does reproduce the $L_x-T$ scaling relations well, it is
ad hoc and fails to produce the observed entropy profiles in a subset of X-ray
luminous, cool core groups with steep entropy profiles and low values of the
central entropy (see Section \ref{sec:intro_groups}). Thus, it is clear that the
effects of heating the ICM are more complicated than the introduction of a
constant, global entropy floor \citep{McCarthy+08,Fang+08}.  Here we explore
whether blazar heating may be capable of producing the observed entropy profiles
as well as some of the features of the distribution of large-scale structures
observed.

While large-scale numerical simulations are required to study the entropy
evolution of groups and clusters in detail, we can use the conservative property
of entropy to estimate the effect of blazar heating on the thermodynamics of
galaxy groups and (to a lesser extent) clusters.  Smooth accretion, in which
cold gas enters the cluster through a spherically symmetric accretion shock,
results in self-similar entropy profiles \citep{Voit2005}.  Characteristic
values for the physical parameters are given by their average values within the
virial radius, $R_{200}$ (that we define as the radius of a sphere enclosing a
mean density that is 200 times the {\em critical} density of the universe):
\begin{equation}
  \begin{aligned}
    n_{e,\,200} &= 200\, x_e X_\rmn{H} f_b \frac{\rho_\rmn{cr}}{m_p}=
    1.7\times10^{-4} E^2(z)\,\cm^{-3}\\
    kT_{200} &= \frac{G M_{200\,c} \mu m_p}{2 R_{200}} = 
    \frac{\mu m_p}{2}\left[10\, G H(z) M_{200\,c}\right]^{2/3}\\
    &= 1\left(\frac{M_{200\,c}\,E(z)}{6\times10^{13}M_\sun}\right)^{2/3}\,\keV \\
    K_{e,\,200} &= \frac{kT_{200}}{n_{e,\,200}^{2/3}} = \frac{\mu m_p^{5/3}}{2}
    \left(\frac{4\pi}{15}\, \frac{G^2 M_{200\,c}}{(1+X_\rmn{H})\,f_b H(z)}\right)^{2/3}\\
    &=
    326\left(\frac{M_{200\,c}\,E^{-1}(z)}{6\times10^{13}M_\sun}\right)^{2/3}\,\keV\,\cm^2\,,
  \end{aligned}
\label{eq:nTK200}
\end{equation}
where $f_b=\Omega_b/\Omega_m$ is the universal baryon fraction, the electron
fraction is defined as the ratio of electron and hydrogen number densities, $x_\e
= n_\e/n_\rmn{H} = (X_\rmn{H}+1)/(2\,X_\rmn{H}) = 1.158$, in which we assumed a
fully ionized fluid with a primordial hydrogen mass fraction $X_\rmn{H}=0.76$,
and $\rho_\mathrm{cr} = \rho_\mathrm{cr} (z)= 3 H^2(z)/(8 \pi G)$ is the
critical mass density, in which the Hubble function $H(z)$ is given by
\begin{equation}
  \label{eq:Hubble}
  \frac{H^2(z)}{H_0^2} = E^2(z) = 
  (1+z)^3 \Omega_m + (1+z)^2 (1-\Omega_m - \Omega_\Lambda) + \Omega_\Lambda\,.
\end{equation}

Of particular importance here is that the entropy scales as $K_e\propto r^{1.1}$
\citep{Voit2005}.  The predictions of this simple model agree well with
numerical simulations \citep{Tozzi+2001} and the entropy profiles in the outer
regions of clusters inferred from recent X-ray observations by {\em Chandra} and
{\em XMM-Newton} \citep{Cavagnolo+2009,Pratt+2010}.  Within the smooth accretion
model the stratified profile is built from the inside out, with later accreted
shells containing larger entropy due to gravitational heating at accretion
shocks.  Thus, if radiative cooling can be neglected, the entropy distribution
of the gas in the cluster core reflects the entropy of the IGM immediately prior
to the initial collapse of the group/cluster. This assumption is reasonable as
long as the radiative cooling time (Equation \eqref{eq:tcool2}) is longer than
the time interval between successive mergers, or equivalently the mass accretion
timescale (see discussion in Sect.~\ref{sec:X-ray}).

In practice, the smooth-accretion picture is over-simplified.  Most of
the ultimately accreted gas is not smoothly distributed, but rather
contained within virialized substructures, and therefore has already
been shock heated, fundamentally altering the way in which the entropy
profile is generated.  This results in a more complex morphology of
the dissipating structure-formation shocks, which exhibit a rich
network of shock fronts at which the gravitational energy of the gas
is dissipated \citep{Miniati+2000,Ryu+2003,Pfrommer+2006}.
Nevertheless, despite this apparent chaos, non-radiative galaxy
cluster simulations that have sufficiently high resolution find
approximately self-similar entropy, density, and temperature
structures outside of the core region, independent of the numerical
method used \citep{Frenk+1999,Voit+2005}, yielding a universal entropy
profile of
\begin{equation}
  \label{eq:Kprofile}
  K_e(r,z) = 1.45\, K_{e,\,200}(z)\,(r/R_{200})^{1.2}.
\end{equation}
The reason for this is that the low entropy/high density gas arranges itself
such that it finds itself at the bottom of the gravitational potential and
eventually mixes with the surrounding halo gas as can be seen by rewriting the
hydrostatic equation,
\begin{equation}
  \label{eq:HSE}
  \frac{dP}{dr} = -\frac{G M(<r)}{r^2} \left(\frac{P}{K}\right)^{3/5}.
\end{equation}
Magnetic draping of cluster fields during gravitational settling provides a
thermal insulation of these low-entropy parcels
\citep{Lyutikov2006,Dursi+2008,Pfrommer+2010}, and hence this settling occurs
adiabatically, producing a stratified core in which the hot, high-entropy halo
gas remains at large radii.

\begin{figure*}
\begin{center}
\includegraphics[width=0.49\textwidth]{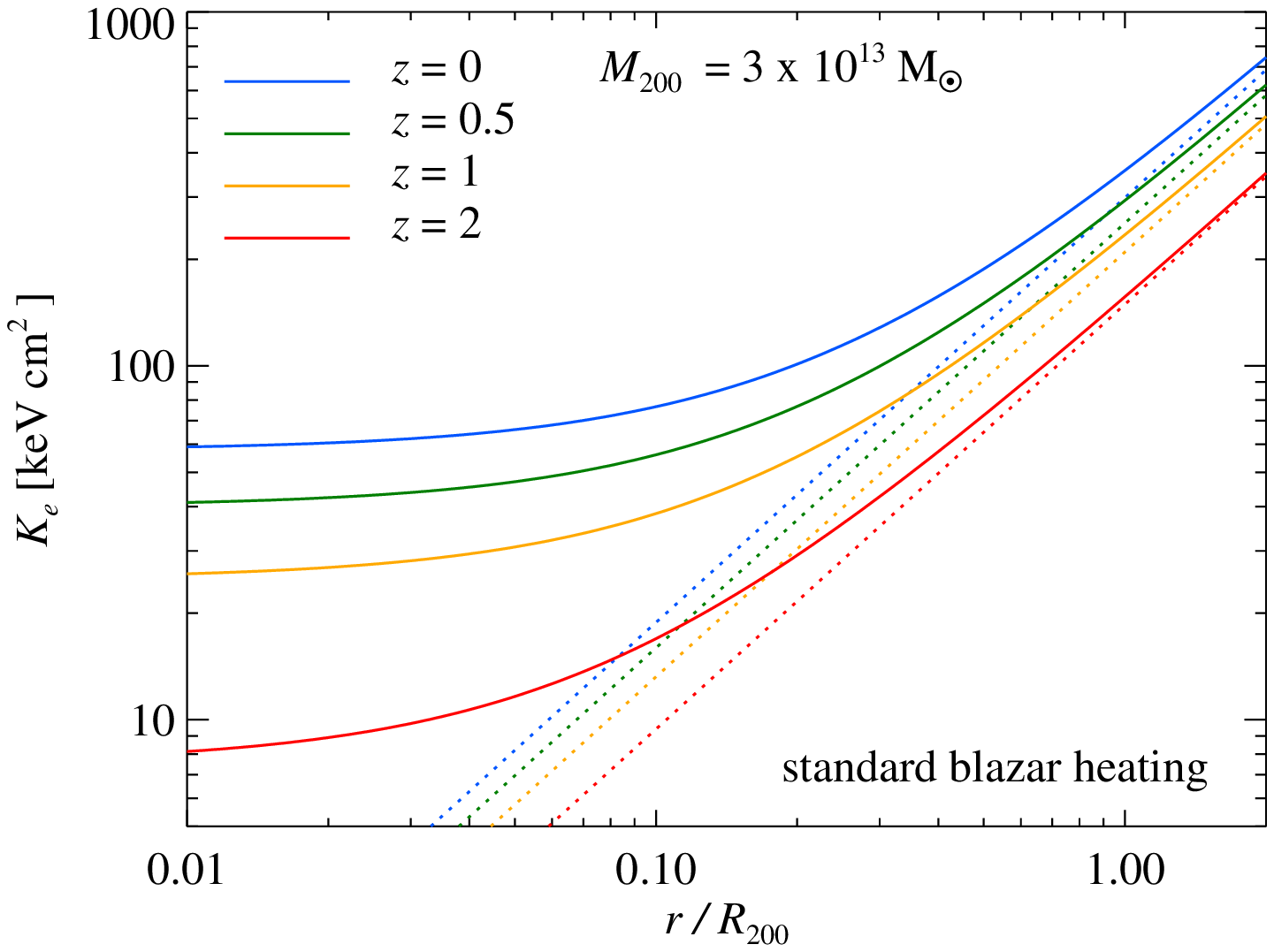}
\includegraphics[width=0.49\textwidth]{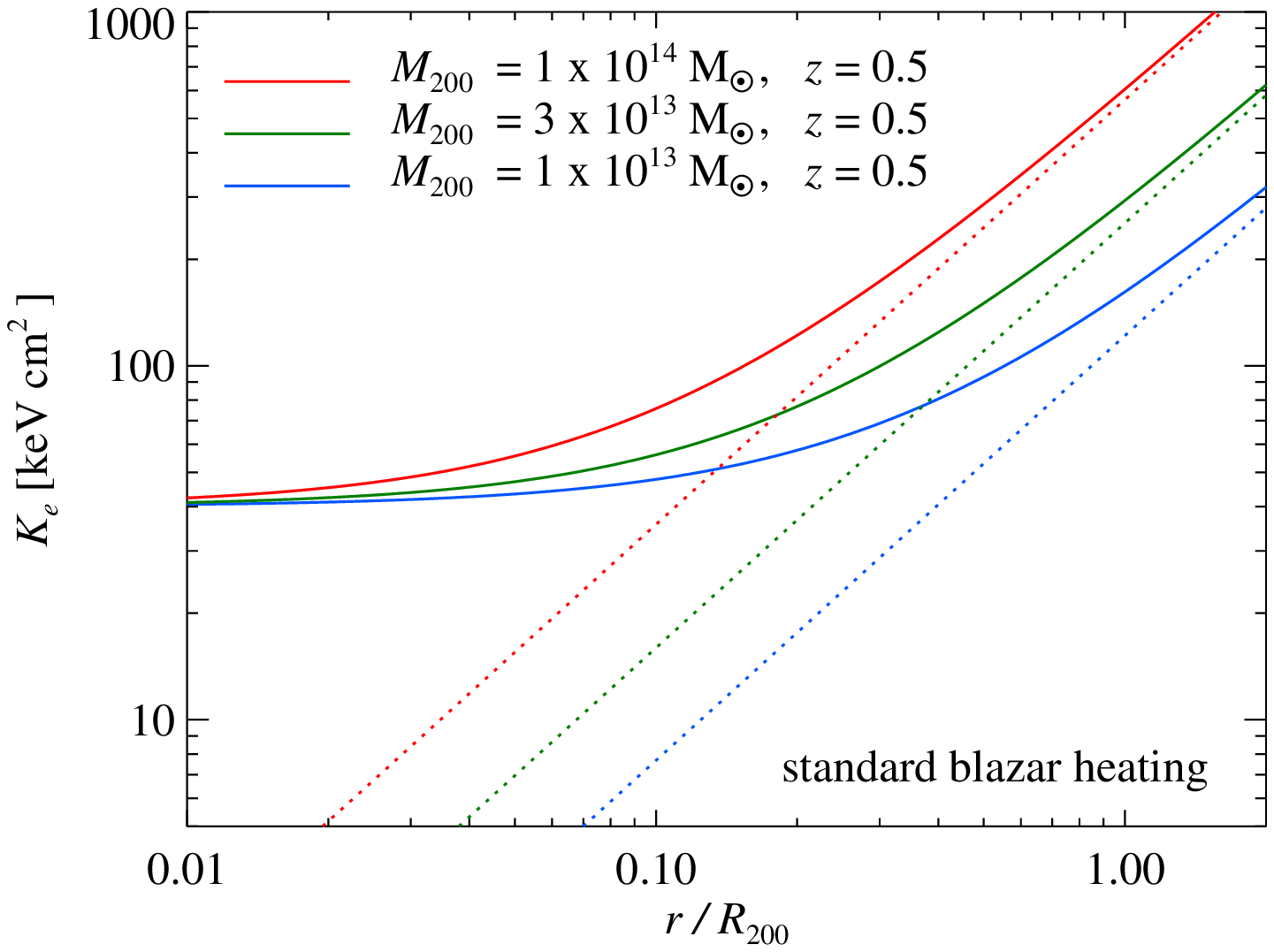}\\
\includegraphics[width=0.49\textwidth]{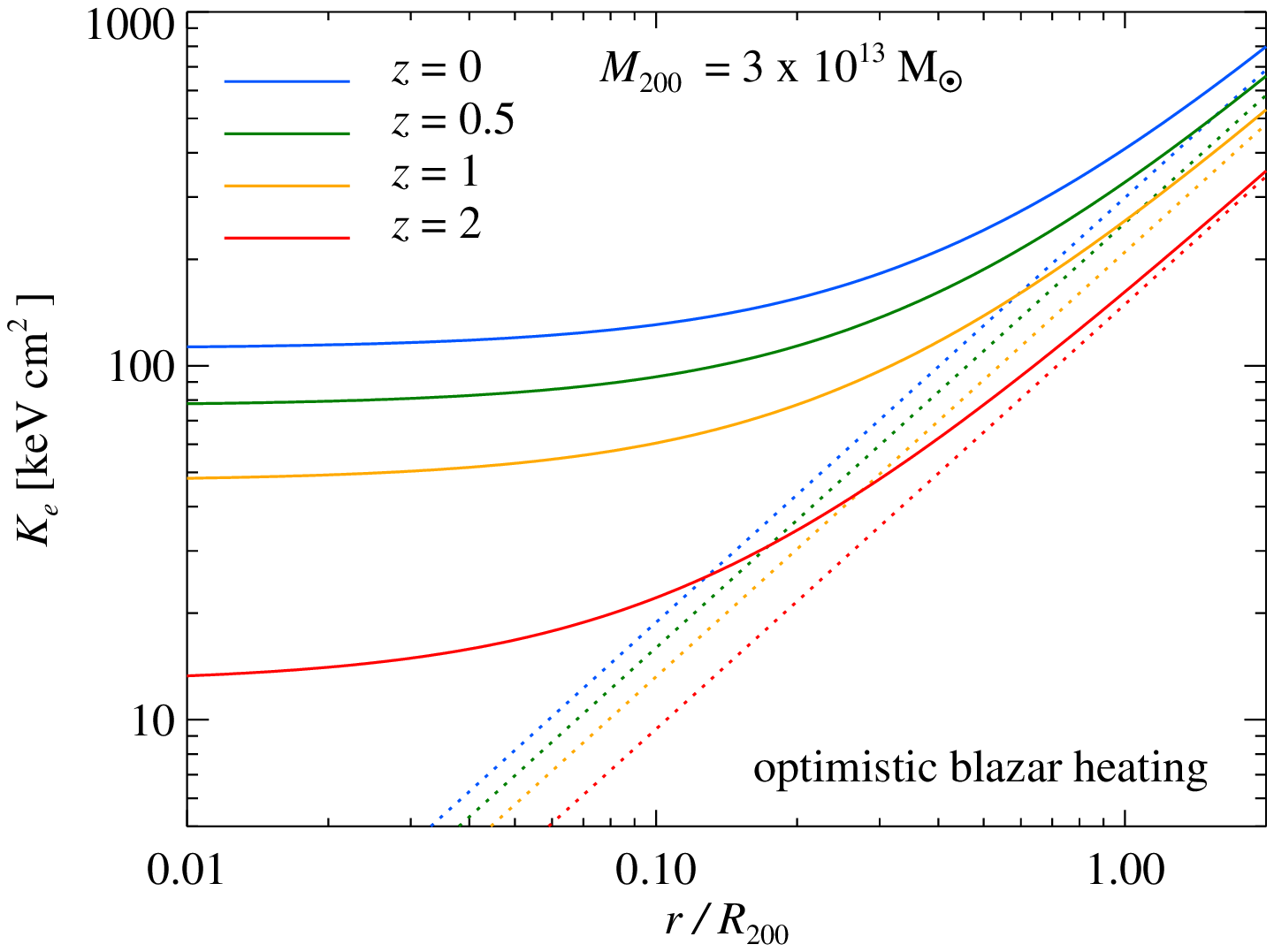}
\includegraphics[width=0.49\textwidth]{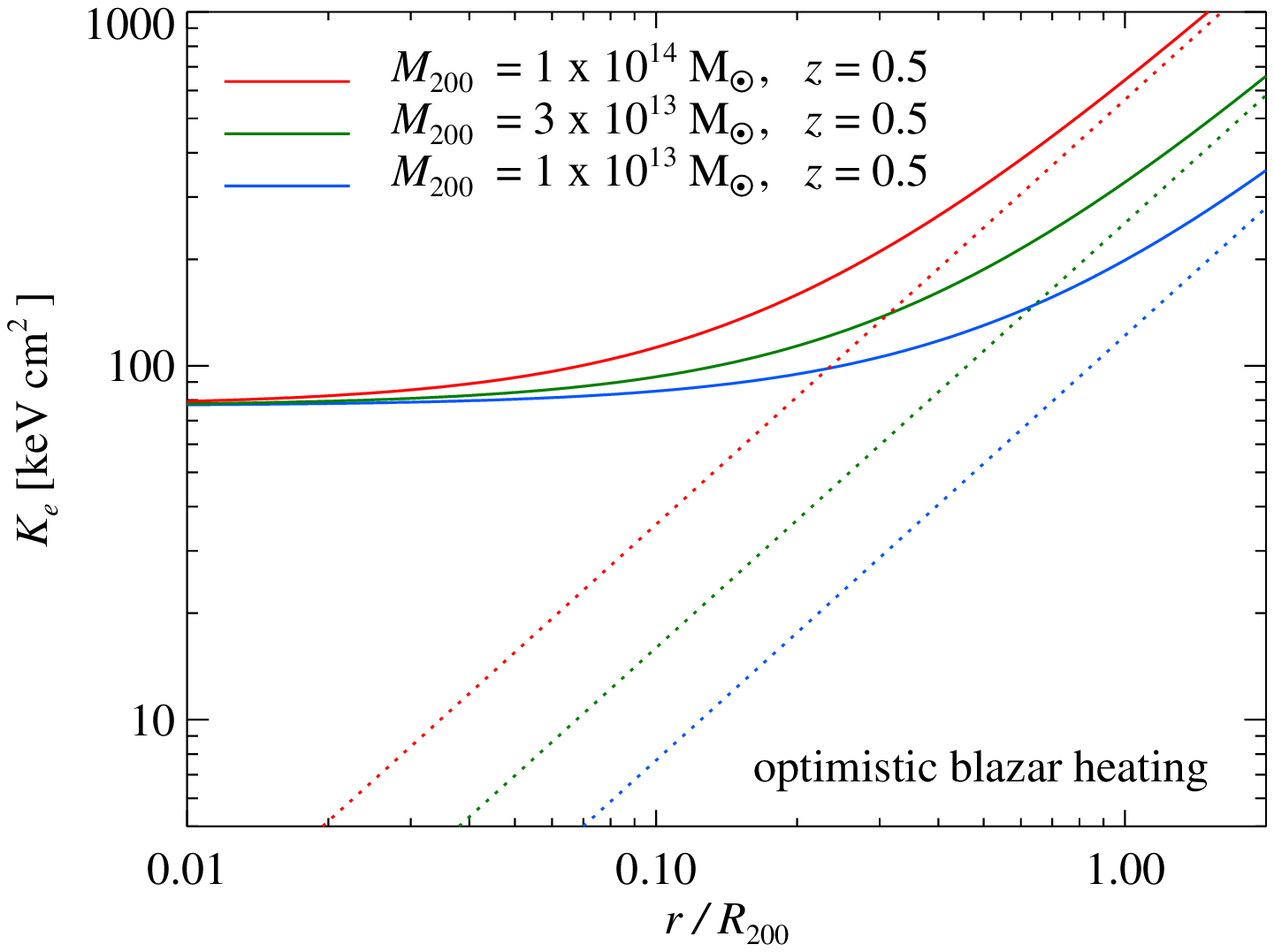}
\end{center}
\caption{Radial profile of the entropy, $K_e = kT / n_e^{2/3}$, in galaxy
  groups/clusters for varying formation redshifts (left) and cluster masses
  (right) when our standard (top) and optimistic (bottom) blazar heating models
  are employed. While the outer profile results from dissipating gravitational
  energy in accretion shocks (dotted), the entropy core immediately after
  formation is set by the redshift-dependent blazar heating. Note that only
  late-forming groups ($z\lesssim1$) are directly affected by this homogeneous
  but redshift-dependent preheating mechanism.  If the ratio of cooling time,
  $t_\rmn{cool}$ to the time period to the successive merger $t_\rmn{merger}$,
  is smaller than unity, the group will radiate away the energy associated with
  the elevated entropy core and evolve into a CC. Alternatively, if
  $t_\rmn{cool}> t_\rmn{merger}$, merger shocks can gravitationally reprocess
  the entropy cores and amplify them (not shown here). Potentially those groups
  may evolve into NCC systems.}
\label{fig:K_r}
\end{figure*}

If the universe is preheated, the central entropy of a newly formed group
or cluster is replaced with a flattened core.  Initially, the entropy of the
core is set by that of the IGM at the time the second gas shell is accreted,
forming an accretion shock and adiabatically compressing the gas in the first.
Employing only heating by blazars and structure formation shocks, we show the
resulting entropy profiles for groups of a variety of virial masses and
formation redshifts in Figure \ref{fig:K_r}, exploring the distribution around
the median object mass.  These are based upon  Equations (\ref{eq:nTK200}) and
(\ref{eq:Kprofile}) as well as the floor values implied by Figure 
\ref{f:thermal history}.  Interestingly, the blazar-induced entropy floor at $z=0.5$ is
comparable to the gravitationally established entropies in groups of
$3\times10^{13}\,M_\sun$ at radii $(0.2$--$0.4)\,R_{200}$ for our standard and
optimistic blazar models, respectively.  The implied core sizes are similar to
observationally accessible radii of $R_{2500}\sim 0.3\, R_{200}$ of optically
selected clusters \citep[see Figure 1 of][]{Pratt+2010}. In our picture, these
correspond to young clusters with recent formation times.

Thus far we have implicitly assumed the instantaneous formation approximation,
i.e., that we may identify a particular redshift at which to calculate
$K_{e,\,200}$ and the core entropy level due to blazar heating.  This is
naturally identified with the redshift at which the group/cluster has assembled
half of its mass, after which strong structure-formation shocks develop.  Figure
\ref{f:macc} shows the accretion histories for groups/clusters in the mass range
of relevance here, noting explicitly the half-mass redshifts for a variety of
accretion histories.  Typically, these occur after $z\simeq1$, by which time
blazar heating has had an opportunity to inject significant amounts of entropy
into the IGM, implying that the TeV blazars can have a substantial impact upon
the structure of groups/clusters in practice.\footnote{The half-mass redshifts
  depend on the adopted definition for the halo mass (and of course cosmology
  which we fix here for simplicity). While we choose to use $M_{200\,c}$ in this
  section, we note that the half-mass redshift are somewhat smaller when
  adopting the definition for the virial mass with an overdensity that varies
  with redshift \citep{Bryan+1998}. In this case, we obtain $z_{0.5} = \{0.85,
  0.74, 0.62, 0.48, 0.36\}$ for our halos, $M_\mathrm{vir} = \{1.3, 3.8, 13, 38,
  130\}\times 10^{13} \,M_\sun$.}  However, we note that median mass accretion
histories are too simplified to assess the impact of blazar heating upon
group/cluster entropy profiles in detail since they do not address the radial
redistribution of accreted gas.  Furthermore, we do not attempt to address the
generation of entropy by gravitational reprocessing during mergers of later
accreted material in Figure~\ref{fig:K_r}. After this explicit demonstration of
the impact of a redshift-dependent entropy floor on cluster entropy profiles, we
turn to the implications of these for the cluster population as a whole.

\subsection{Implications for the bimodality of core entropy values}\label{sec:bimod}

\begin{figure*}
\begin{center}
\includegraphics[width=0.495\textwidth]{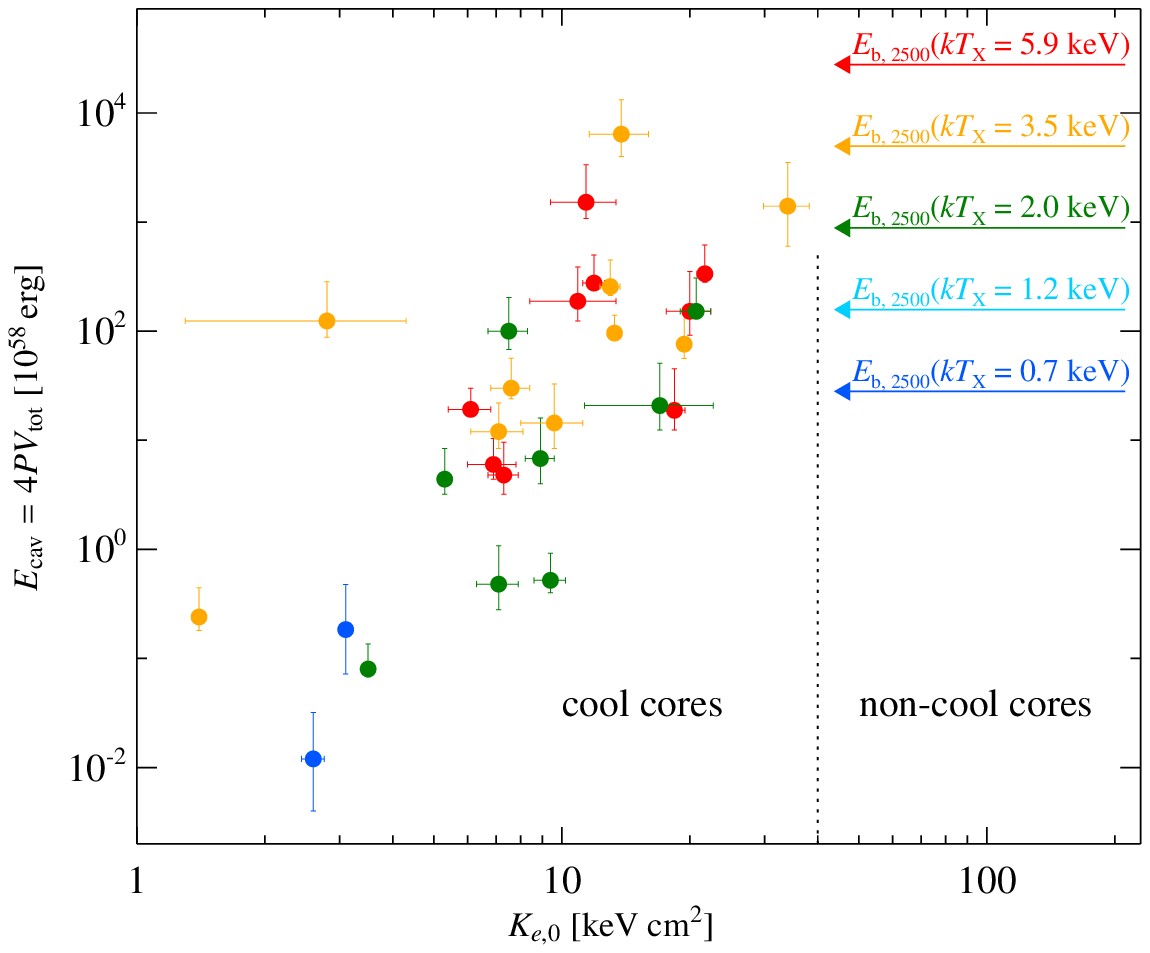}
\includegraphics[width=0.495\textwidth]{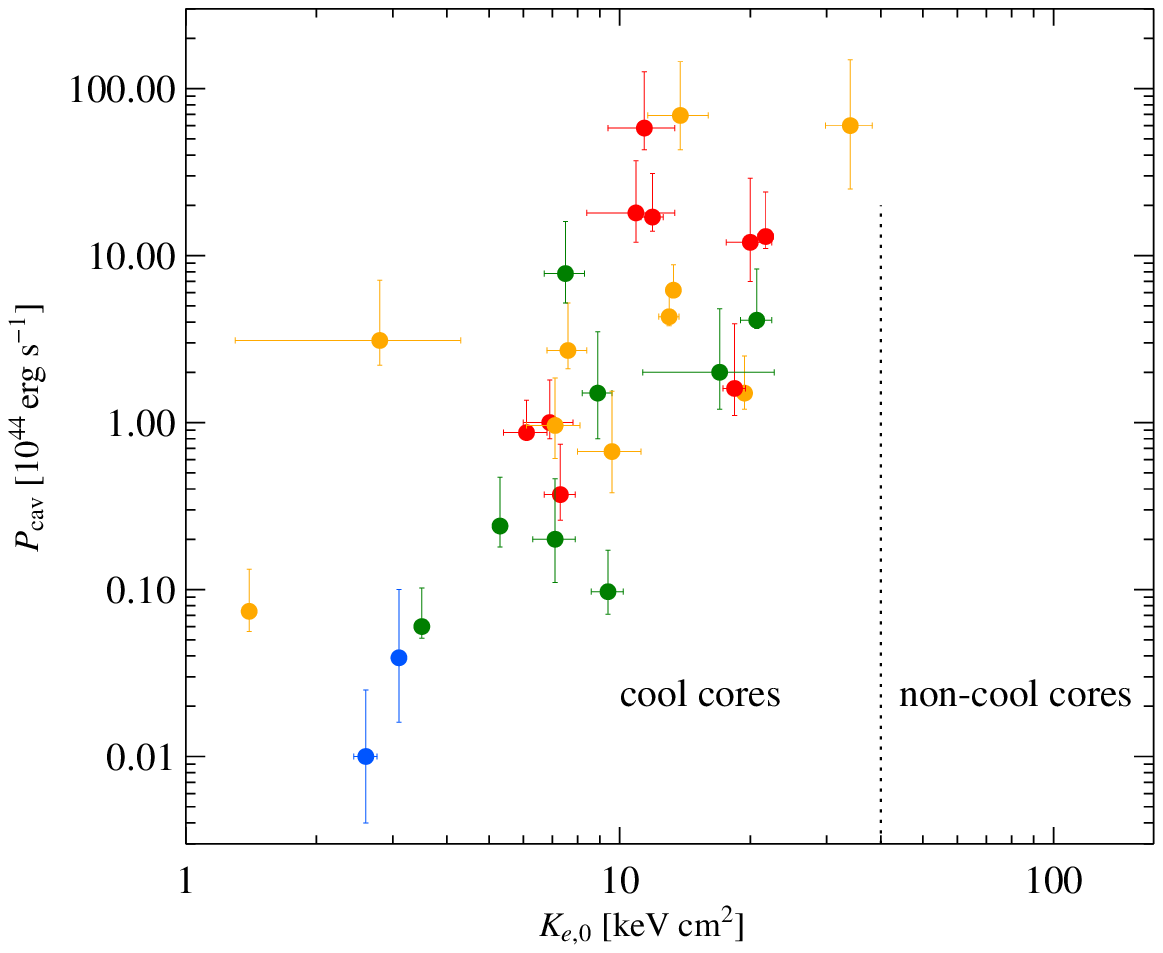}
\end{center}
\caption{Correlating the observed core entropy values of clusters, $K_{e,0}$,
  \citep[taken from the gradient extrapolation method of][]{Cavagnolo+2009} with
  the AGN cavity energy $E_\rmn{cav}$, as inferred from the volume work done by
  the expanding bubbles (left), and with the cavity power
  $P_\rmn{cav}=E_\rmn{cav}/t_\rmn{buoyancy}$ \citep[right,
  from][]{Rafferty+2006}. Color coding reflects average X-ray temperatures, the
  lower limit of each color bin is labeled in the upper right of the left
  panel. Arrows denote the gas binding energy contained within a spherical
  region of radius $R_{2500}\simeq R_{200}/3$.  These very powerful AGN
  outbursts with energies up to $10^{62}\,\erg$ and powers of
  $10^{46}\,\erg\,\s^{-1}$ are in some cases energetically capable of unbinding
  the gas in the core regions.  However, the mechanical energy of these
  expanding cavities only heat the cluster core enough to prevent a cooling
  catastrophe.  On the buoyancy timescale, no AGN outburst transforms a CC to a
  NCC cluster.  This is apparent from the low core entropy values (with a median
  $K_{e,0} = 15\,{\rm keV\,cm^2}$) of typical CC clusters.}
   \label{f:agn}
\end{figure*}

The thermodynamic properties of clusters in their centers show a clear
bimodality, which is traditionally separated into two classes: CC and NCC
clusters. The former are defined to have temperature profiles which decline
significantly toward the center whereas the central temperature distribution of
the latter remains constant and often correlates with merger events. As a
result, the distribution of core entropy also appears to be bimodal in clusters,
with the CC population peaking at $K_{e,0} \sim 15 \,\keV\,\cm^2$ and NCCs at
$K_{e,0} \sim 150 \,\keV\,\cm^2$ separated by a gap between $K_{e,0} \sim
(30$--$50) \,\keV\,\cm^2$.  Roughly half of the entire population of galaxy
clusters in the {\em Chandra} archival sample show cooling times that are longer
than 2~Gyr and have high-entropy cores typical of NCCs \citep{Cavagnolo+2009}.
This CC/NCC bimodality appears to be real and not due to archival bias as a
complementary approach has shown with a statistical sample \citep{Sanderson+2009}.

Before we discuss how blazars may impact on the relative abundance of cluster
populations, we will shortly review how the currently favored hypothesis of AGN
feedback compares to data.  The possibility that AGN feedback can raise the core
entropy $K_{e,0}$ to values representative of NCCs on the buoyancy timescale
\citep[e.g.,][]{Guo+2009} is not supported by observations. This is explicitly
shown in Figure~\ref{f:agn} which correlates $K_{e,0}$ with the volume work done
by the expanding bubbles, $E_\rmn{cav}$, and with the cavity power,
$P_\rmn{cav}$, in systems that show X-ray cavities inflated by AGN bubbles.  The
total energy required to inflate a cavity is equal to its enthalpy, given by
\begin{equation}
  \label{eq:Ecav}
  E_\rmn{cav} = \frac{\gamma}{\gamma-1}\,P V_\rmn{tot} = 4\,P V_\rmn{tot} 
\end{equation}
assuming a relativistic equation of state $\gamma=4/3$ within the bubbles. The
cavity power is estimated using $P_\rmn{cav}=E_\rmn{cav}/t_\rmn{buoyancy}$
\citep{Birzan+2004, Rafferty+2006}.\footnote{The buoyancy timescale could be
  either an overestimate of the true age since the cavity is expected to move
  outward supersonically during the early, momentum-dominated phase of the
  jet. It could also be an underestimate of the true age since magnetic draping
  provides an additional drag force slowing down the rise of the bubble
  \citep{Dursi+2008}.} We compare the energy used to inflate the cavities to the
gas binding energy of the core region within a spherical region of radius
$R_{2500}\simeq R_{200}/3$,
\begin{equation}
  \begin{aligned}
    E_{b,2500} &= f_\rmn{gas,2500}\,\frac{G M_{2500}^2}{2 R_{2500}} = 
    \frac{f_\rmn{gas,2500}}{2} M_{2500}^{5/3}\left[10\, G H(z) \right]^{2/3}\\
    &\simeq 1\times 10^{60}\,\erg \left(\frac{k T_X}{1\,\keV}\right)^{3.23}
  \end{aligned}
\label{eq:Ebind}
\end{equation}
at $z=0$. Here we use the phenomenological scalings obtained by X-ray
observations of $h_{70}\, M_{2500} = M_5\, (kT_\rmn{X} / 5\,\keV)^{1.64\pm0.06}$, with
$M_5 = (2.5\pm0.1)\times 10^{14}\, h_{70}^{-1} \,\rmn{M}_\sun$ \citep{Vikhlinin+2006}
and $f_{\rmn{gas},2500} = (0.0347\pm0.0016)\, (kT_\rmn{X} / 1\,\keV)^{0.509\pm0.034}$
\citep{Sun+2009}.

$P_\rmn{cav}$ or $E_\rmn{cav}$ both measure the energy and power that is in
principle available for heating the ICM, if it can be efficiently tapped by some
process.\footnote{We note that of this $4\,P V_\rmn{tot}$, only $P V_\rmn{tot}$
  is available in form of mechanical energy while the internal energy $U=3\,P
  V_\rmn{tot}$ is presumably still stored within the bubbles.  What fraction of
  this internal energy is eventually thermalized, and thus can potentially
  contribute to unbinding the cluster gas and/or raising the core entropy, is
  not a priori clear. If the energy is stored in cosmic rays or magnetic fields,
  it may be transferred to the thermal pool via cosmic ray Alfv\'en-wave heating
  \citep{Kulsrud+1969} or magnetic reconnection, respectively.  Nevertheless,
  the uncertainty induced by this is small in comparison to the many orders of
  magnitude increase in cavity energy required to explain the non-cool core
  clusters.} However, as shown in Figure~\ref{f:agn}, even the most energetic
and powerful AGN outbursts, with $E_\rmn{cav} \sim 10^{62}\,\erg$ and
$P_\rmn{cav} \sim 10^{46}\,\erg\,\s^{-1}$, e.g., MS0735+7421 and Zwicky 2701,
which are energetically capable of unbinding the gas in the core regions, are
unable to disrupt the CC and transform the cluster into an NCC state on a
buoyancy timescale.  This is apparent from the low core entropy values (with a
median $K_{e,0} = 15\,{\rm keV\,cm^2}$) of typical CC clusters.  This is
quantified with a linear Pearson correlation coefficient of 0.71 correlating
$\log K_{e,0}$ with $\log P_\rmn{cav}$ or $\log E_\rmn{cav}$, respectively.
This proves that the $PdV$ work done by these expanding cavities is transferred
inefficiently to the surrounding medium, i.e., the ICM entropy produced by AGN
inflated bubbles is far less that the virial value of $K_e \sim
540\,\keV\,\cm^2\times (M_{200\,c}/\rmn{M}_{14})^{2/3}$, but enough to arrest
overcooling.

On the other hand, there is a strong anticorrelation between the radio power of
the brightest cluster galaxy and $K_{e,0}$ for nearby clusters ($z< 0.2$),
implying that bright radio emission is preferentially ``on'' for $K_{e,0}
\lesssim 40\,\keV \cm^2$ \cite[see Figure 2 in][]{Cavagnolo+2008}. While AGN
feedback seems to be unable to transform a CC to an NCC cluster (on a buoyancy
timescale), it appears to be critical in stabilizing the thermal atmospheres
from entering a cooling catastrophe and collapsing. In principle the impact of
AGN-induced turbulence on heat transport (conductively and advectively) could
result in a CC to NCC metamorphosis on a much longer ($>$ Gyr) timescale
\citep{Parrish+2010,Ruszkowski+2010}. This is because the temperature difference
between the maximum of the temperature profiles to the cold center is at most a
factor of three \citep{Vikhlinin+2006}, implying that heat transport could
initially increase the central entropy by a similar factor as $K_{e,0} \propto
kT$. In the absence of radiative cooling, the associated pressure enhancement
would adiabatically expand and thereby cool the gas, hence restoring the
temperature gradient. Sustained conduction and advection could further increase
the central entropy. However, the timescale for such a process is long in
comparison to radiative cooling timescales ($\lesssim 1$ Gyr) as well as
cluster assembling and merging timescales which questions the possibility of
this mechanism to boost the central entropies to values of $K_{e,0} \simeq
600\,\keV\,\cm^2$ which are at the tail end of the distribution
\citep{Cavagnolo+2009}.

Because blazar heating does not inject large amounts of entropy at $z\gtrsim2$,
and blazars do not efficiently heat high density regions ($1+\delta>10$),
objects that have already collapsed by that time or shortly thereafter will not
be significantly affected by blazars (barring the effect of clustering bias at
early redshifts). In our scenario, these early-forming groups evolve into CC
systems at the present epoch that potentially need to be stabilized by a
self-regulated feedback process, e.g., provided by the radio mode of AGNs.
However, the subset of groups that forms after $z\simeq1$ can be severely
affected. If such a group is viewed shortly after its formation, it should still
exhibit the elevated core entropies associated with the prior blazar heating.
If such a late-forming group has a cooling timescale long in comparison to the
interval between cluster/group mergers, merger shocks can gravitationally
reprocess the entropy cores and amplify them by a factor of up to five
\citep[][see also Section \ref{sec:X-ray}]{Borgani+2005}.  These late-forming
groups would then evolve into NCC systems.\footnote{Using {\em Chandra}
  observations, there has been a claim that incidence rate of cool core clusters
  at redshifts $z>0.5$ is much smaller than their fraction at low redshifts
  \citep{Vikhlinin+2007}. We note that the development of the characteristic
  cuspy X-ray brightness profiles for cool cores at $z=0$ requires the cooling
  time to be considerably shorter than the formation time---a criterion that is
  often not fulfilled at the (high) redshift in question. Second, in order to
  construct their sample, \citet{Vikhlinin+2007} discarded regions with AGN
  emission which can be correlated with cool cores, potentially biasing their
  absolute cool core rates low. Hence, this observation is not in contradiction
  with our proposed scenario.}  For sufficiently late-forming clusters that
experience a series of fast successive merger events and, hence, avoid
substantial cooling phases, we expect that gravitational reprocessing should
boost the central core entropy in the most extreme case from
$\sim100\,\keV\,\cm^2$ to $\sim600\,\keV\,\cm^2$, allowing for reprocessing of
the blazar-heated entropy floor due to gravitational heating. Typically,
however, blazar heated entropies at the time of turnaround for late-forming
groups/clusters are $\sim 50\,\keV\,\cm^2$. Allowing for modest cooling periods
in between mergers should yield smaller median core entropy values of
$\sim(100-150)\,\keV\,\cm^2$ after accounting for gravitational
reprocessing. These estimates compare favorably with observed values of NCC
clusters \citep{Cavagnolo+2009}.

We point out that it is very natural for systems with a unimodal
distribution in core entropy values following group formation to
evolve into a bimodal distribution today as consequence of both the
cooling instability and the gravitational reprocessing of (temporally
increasing) elevated entropy cores.  The observed CC and NCC cluster
populations are centered upon the two attractor solutions that a
galaxy cluster can evolve into.  In the 
case of CC clusters, the core evolution is driven by the well-known overcooling
problem, encountered in cosmological simulations of galaxy clusters.  Below a
critical core entropy, purely hydrodynamical mergers are incapable of disrupting
a compact CC systems and transforming it into an NCC object \citep{Poole+2008}.
For NCC clusters, the core evolution is driven by the rapid (in comparison to
$t_\rmn{cool}$, which effectively sets the critical core entropy value within a
given epoch) succession of mergers, which has been demonstrated to further
elevate the core entropy values substantially \citep{Borgani+2005}.
Interestingly, this solution is not a runaway solution; instead, the
gravitational bootstrapping of blazar preheated entropy adjusts to the system
size.  Hence, in this picture, the core entropy should never be able to exceed
the entropy at the virial radius (according to virial arguments) and most likely
reach only values that are a fraction of that (at least for cluster systems) due
to radiative cooling and the modest blazar preheated entropy values in
comparison to early preheating models.

To conclude, we demonstrate explicitly that the core entropy values in CC
clusters have a very weak correlation with the mechanical energy and power of
X-ray cavities inflated by AGNs. Hence, $PdV$ work done by these expanding
cavities is transferred inefficiently to the surrounding medium.  This strongly
suggests that while AGN feedback seems to be critical in stabilizing CC systems,
it cannot transform CC into NCC systems (at least on the buoyancy
timescale). With this evidence it seems even more pressing to pursue alternative
solutions such as the presented blazar heating scenario in combination with
gravitational reprocessing that provides a plausible scenario for the observed
CC/NCC bimodality. Future cosmological hydrodynamical simulations that include
the effect of clustering bias of blazar heating are needed to study these
considerations in greater detail.\footnote{A detailed prediction of the core
  entropy distribution of clusters at a given mass (or temperature) depends on
  the merger history of all objects at that mass (to understand which number
  fraction of clusters was channeled into the cooling branch due to
  comparatively fast radiative cooling), the distribution of the departure times
  of the gas that ended up in the core from average densities (to address the
  exact magnitude of blazar heating for the population of clusters rather than
  individual systems), and quantifying the efficiency of gravitational
  re-processing for the entire population of groups and clusters
  (\citet{Borgani+2005} only simulated a group and a cluster for different
  variants of physics).}

\subsection{Implications for the Sunyaev-Zel'dovich power spectrum}\label{sec:SZ}

The thermal SZ effect provides a direct probe of the gas properties of groups
for $z \gtrsim 0.5$.  The SZ effect arises from CMB photons that inverse Compton
scatter off thermal electrons within the hot plasma in galaxy clusters and
groups, producing a localized perturbation to the CMB spectrum
\citep{SZ1972,SZ1980}.  The thermal SZ effect directly measures the thermal
electron pressure in the gas and has the important property that its amplitude
is independent of redshift.  The pressure fluctuation spectrum of unresolved
groups and clusters dominates the CMB power spectrum on angular scales smaller
than $3\arcmin$ (corresponding to a multipole moment $\ell\simeq3000$) and half
of the SZ power spectrum signal at $\ell\simeq3000$ comes from groups with
$M_{500}<2\times10^{14}M_\sun$ and $z>0.5$ \citep{Trac+2011,Battaglia+2011}. At
these scales, the SZ power spectrum depends on the square of the Fourier
transform of the average pressure profile of clusters/groups; a more
concentrated pressure profile implies more power, a smoother one less.

A population of groups/clusters with high core entropies,
$K_0=(50$--$100)\,\keV\,\cm^2$, implies a smoother pressure core distribution
than if the cores had cooled since formation and developed a more concentrated
entropy profile.  This is very similar to the effect of AGN feedback which
injects entropy into the core of groups, smoothing out the resulting pressure
profile \citep{Battaglia+2010}.  Adopting empirically motivated ``universal''
pressure profiles constrained by X-ray observations, the peak amplitude of the
SZ power spectrum is reduced for NCC pressure profiles as compared to those of
CC clusters \citep{Efstathiou+2011}.  Hence, we expect the effect of blazar
heating to result in a suppression of the SZ power spectrum for scales
$\ell\gtrsim2000$, which probe the pressure profile mostly inside $R_{500}$. An
abundant population of late-forming groups with high $K_0$-values (that has not
been taken into account in any numerical modeling of the SZ power spectrum so
far, e.g., by \citealt{Battaglia+2010} or \citealt{Trac+2011}) would furthermore
reduce the thermal SZ power spectrum in comparison to these numerical
approaches. This has potentially important observational consequences since
angular scales around $3\arcmin$ are the sweet spot for current telescopes
measuring the high-$\ell$ CMB angular power spectrum, e.g., the South Pole
Telescope
\citep[SPT;][]{Lueker+2010,Shirokoff+2010,Keisler+2011,Vanderlinde+2010} and the
Atacama Cosmology Telescope \citep[ACT; see,
e.g.,][]{Fowler+2010,Dunkley+2010,Marriage+2010}.

In addition to the astrophysical dependence, the amplitude of the SZ power
spectrum also depends very sensitively on cosmological parameters that are
responsible for the growth of structure, $C_\ell\propto \sigma_8^{7\ldots9}
(\Omega_b h)^2$, where the rms amplitude of the (linear) density power spectrum
on cluster-mass scales is denoted by $\sigma_8$. Hence, there is an interesting
degeneracy in the amplitude (and shape) of the SZ power spectrum between the
cosmological information (dominated by $\sigma_8$) and the astrophysical
information contained in the average pressure profile.  Currently, numerical
models of the SZ power spectrum are consistent with the data at the 1 $\sigma$
level \citep{Dunkley+2010,Shirokoff+2010}.  However, after allowing for a
substantial signal from patchy reionization \citep{Iliev+2007,Iliev+2008}, which
boosts the kinetic SZ effect and hence the total SZ signal, the power predicted
by these models becomes uncomfortably high. Suppression of power due to blazar
heating represents a promising mechanism by which to reconcile the expected and
observed SZ power spectrum.  

Unfortunately, the impact of blazar heating is degenerate with energy injection
from AGN feedback in the dense cores of groups at early times (where there are
also very little constraints). Here, we sketch a promising idea on how to
discriminate between the effects of AGN feedback and blazar heating on the SZ
power spectrum. In particular, AGN feedback and blazar heating vary as a
function of cluster mass and redshift. Their effects on the pressure profile of
groups and clusters and, hence, their imprint on the SZ signal also vary as a
function of cluster mass and redshift. By cross-correlating CMB maps (where
foreground components and primary anisotropies have been subtracted) with deep
optical redshift surveys that are separately binned in redshift and cluster
mass, i.e., optical richness estimators, we can perform SZ tomography.  Such
tomographic SZ power spectra \citep[similar to Figures 7 and 8 of][for the case
of AGN feedback]{Battaglia+2011} will enable us to derive a redshift and
mass-dependent mean cluster pressure profile. Potentially, this could
disentangle the different non-gravitational energy injections of AGN feedback
and blazar heating in clusters. However, large cosmological hydrodynamical
simulations are required to obtain detailed predictions of either process which
shall be subject to future work.

\section{Structure formation and dwarfs}\label{sec:structure formation}

The entropy injected into the IGM by blazars not only modifies the structures of
groups and potentially clusters, but also has an observable effect upon
cosmological structure formation.  Heating the IGM produces higher IGM
pressures, which in turn suppress the gravitational instability on sufficiently
small scales.  Thus, there is a characteristic length scale, and hence
characteristic mass ($M_C$), below which objects will not form.  The particular
value of this critical mass depends upon how structures form in practice, and
generally requires a fully nonlinear study of structure formation (e.g., that
provided by large-scale numerical simulations).  Nevertheless, we may estimate
the relevant characteristic length and mass scales via linear perturbation
theory.

A very rough idea of the impact of blazar heating may be derived from the
Jeans wavenumber, obtained by balancing the sound crossing and free-fall
timescales:
\begin{equation}
  \label{eq:Jeans_scale}
  k_J(a) \equiv \frac{a}{c_s(a)}\,\sqrt{4 \pi G \bar{\rho}(a)}.
\end{equation}
where for convenience, we have introduced the cosmic scale factor,
$a\equiv1/(1+z)$, $\bar{\rho}(a)\equiv\Omega_m(a)\rho_\mathrm{cr} (a)$ is the
mean total mass density of the universe, $c_s(a)\equiv \sqrt{5 k T(a)/3 \mu
  m_p}$ is the linear sound speed (in which $T$ denotes the temperature at mean
density and $\mu=0.588$ denotes the mean molecular weight for a fully ionized
medium of primordial element abundance).  In a static background universe,
perturbations on scales smaller than $2\pi a/k_J$ are stable, i.e., the ambient
gas pressure is sufficient to counteract gravitational collapse.  As a result,
cooling, fragmentation, and star formation are suppressed within objects of mass
less than the Jeans mass, $M_J$,
\begin{equation}
  \label{eq:Jeans_mass}
  M_J(a) \equiv \frac{4\pi}{3}\,\bar{\rho}(a)\,\left(\frac{2\pi a}{k_J(a)}\right)^3 = 
  \frac{4\pi^{5/2}}{3}\,\frac{c_s^3(a)}{G^{3/2}\bar{\rho}^{1/2}(a)}\,.
\end{equation}

The Jeans mass associated with a universe which has been heated by TeV
blazars, $M_{J,{\rm blazar}}$, is necessarily larger than one which has
only undergone photoionization heating, $M_{J, {\rm photo}}$ by
\begin{equation}
  \label{eq:Jeans_comp}
  \frac{M_{J, {\rm blazar}}}{M_{J, {\rm photo}}} = 
  \left(\frac{c_\mathrm{s,blazar}}{c_\mathrm{s,photo}}\right)^3 =
  \left(\frac{T_\mathrm{blazar}}{T_\mathrm{photo}}\right)^{3/2} 
  \gtrsim \left\{
\begin{array}{rl}
18, & \mbox{stand. model}, \\
50, & \mbox{opt. model}, 
\end{array}
\right.
\end{equation}
where we used a temperature ratio of
$T_\mathrm{blazar}/T_\mathrm{photo}\gtrsim\{7, 14\}$ at the present
epoch for the standard and optimistic model, respectively (see
Figure~\ref{f:thermal history} and the discussion surrounding the
standard and optimistic blazar heating histories).  That is, blazar
heating increases the mass of the smallest collapsed objects by more
than an order of magnitude with the exact value depending on the
adopted numbers of blazars contributing to the heating.

While the Jeans mass provides a way to estimate the importance of blazar
heating, $M_J$ typically exceeds $M_C$ by up to an order of magnitude
because it neglects the Hubble expansion.  That is, it fails to account for the
time required for the pressure to influence the evolving gas distribution.  A
more rigorous treatment of linearized density perturbations in a baryon-dark
matter fluid finds that the overdensity in dark matter, $\delta_d(t,k)$, and the
overdensity in the baryon distribution, $\delta_b(t,k)$, both of which are
functions of time and comoving wavenumber, are related by
\begin{equation}
  \label{eq:filtering}
  \frac{\delta_b(t,k)}{\delta_d(t,k)} = 1 - \frac{k^2}{k_F^2} + \mathcal{O}(k^4)\,,
\end{equation}
for some $k_F$ \citep{Gnedin+1998}.  The ``filtering scale'',
associated with $k_F$, defines a size below which baryonic
perturbations are smoothed despite the growth of background dark
matter perturbations.  The filtering wavenumber can be related to
$k_J$ by
\begin{multline}
  \frac{1}{k_F^2(t)} = \frac{1}{D_+(t)}\int_0^t dt' a^2(t') 
  \frac{\ddot{D}_+(t') + 2 H(t')\dot{D}_+(t')}{k_J^2(t')}
  \int_{t'}^t \frac{dt''}{a^2(t'')}\,,
  \label{eq:Jeans_Dplus}
\end{multline}
where $D_+(t)$ is the linear growth function and is dependent upon the cosmology
\citep{Gnedin2000}.  While $k_F$ is related to $k_J$, at any time those
can be very different since $k_F$ is an integral over the past evolution of the
Jeans scale, weighted by the appropriately scaled growth function.

\begin{figure*}
\begin{center}
\includegraphics[width=0.495\textwidth]{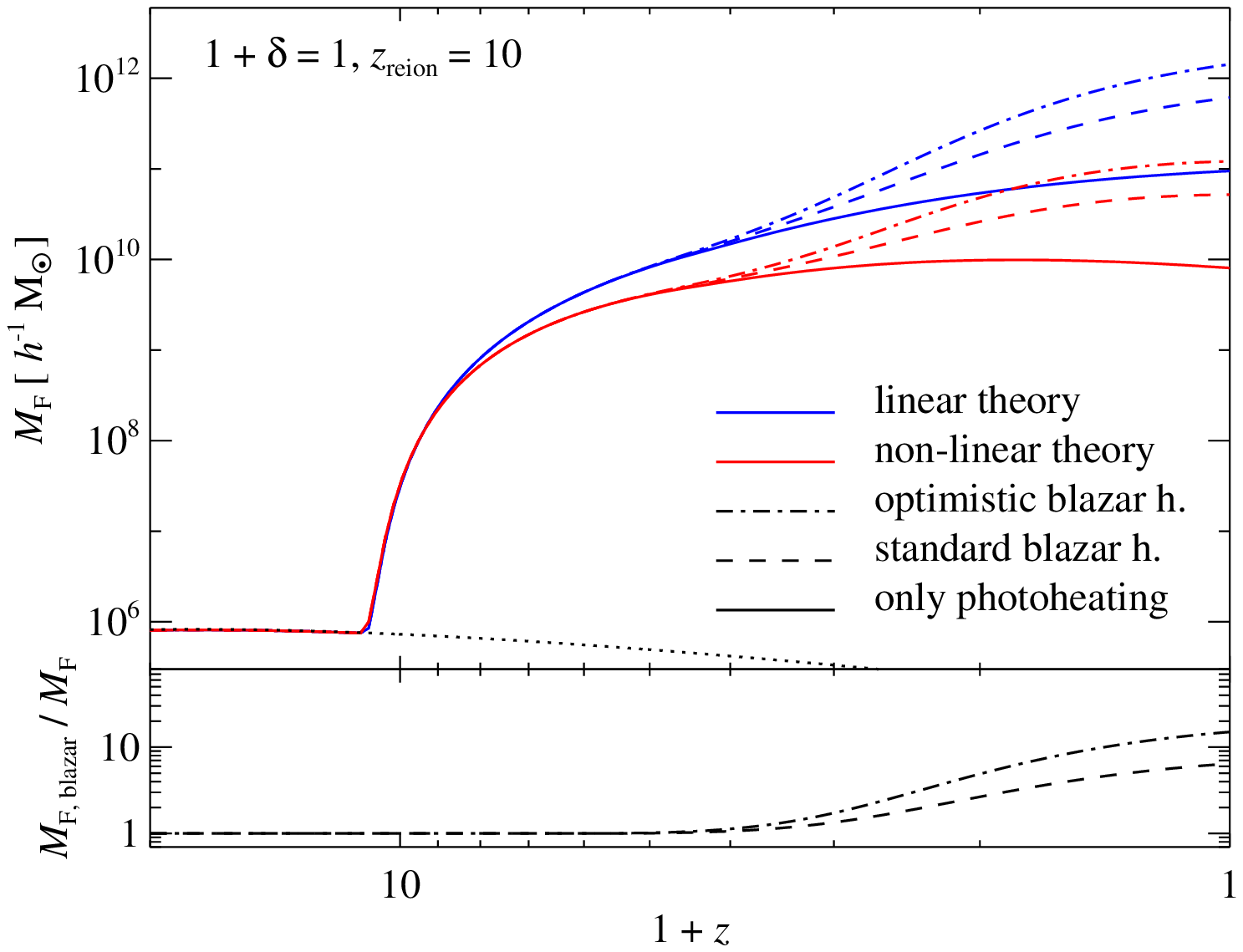}
\includegraphics[width=0.495\textwidth]{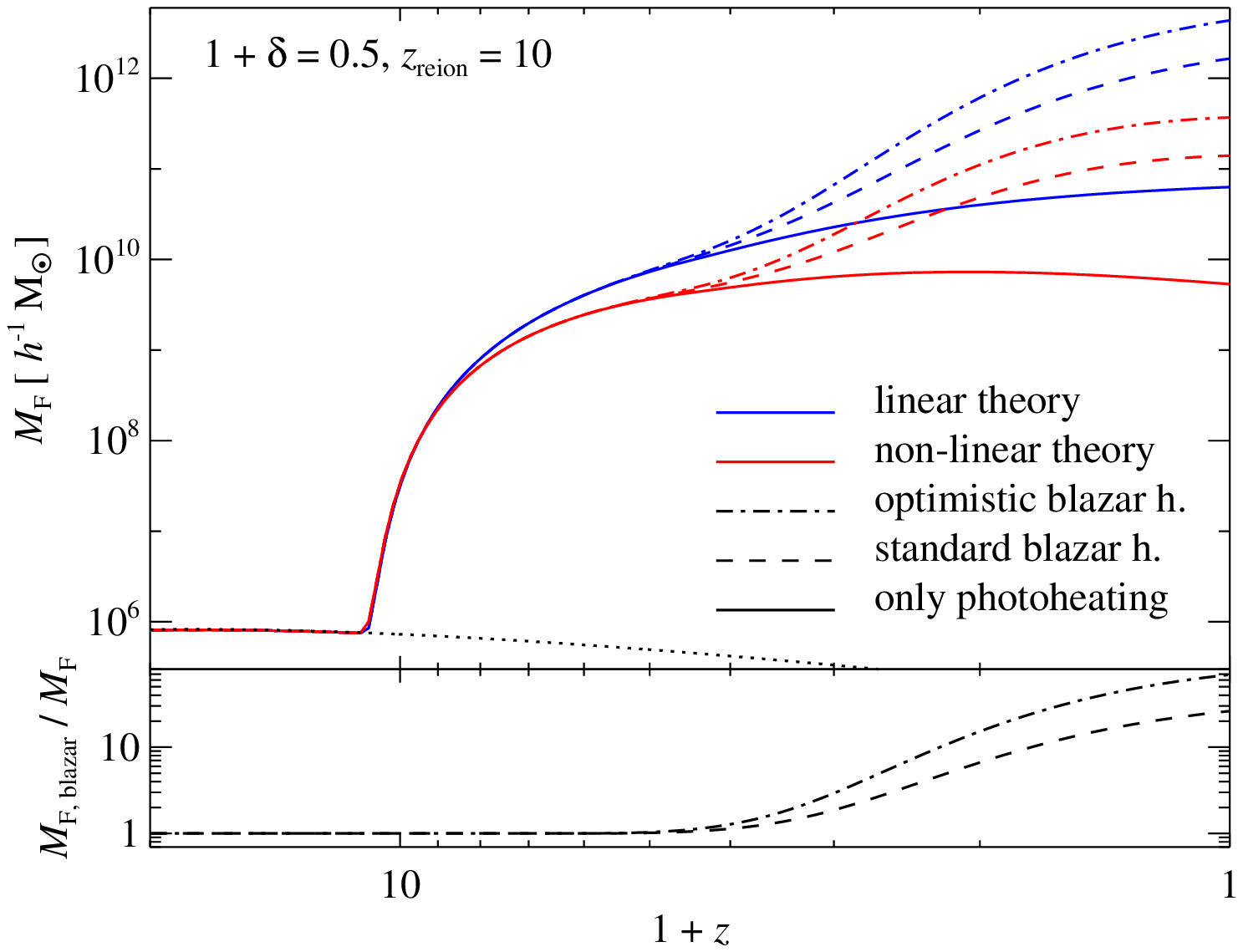}
\end{center}
\caption{Redshift evolution of the filtering mass, $M_F$, for the cosmic
  mean density, $\delta=0$ (left) and for a void with mean overdensity,
  $\delta=-0.5$, (right). We contrast $M_F$ in the standard cosmology that
  employs only photoheating (solid) to the case of blazar heating in our
  standard model (dashed) and optimistic model (dash-dotted). In the bottom
  panels, we show the ratio of $M_F$ in our respective blazar heating models to
  those without blazars.  To estimate the effect of nonlinear structure
  formation on the filtering mass, we compare the linear theory $M_F$ (blue) to
  the nonlinear theory $M_F$ (red) where we used a correction function derived
  from hydrodynamic simulations by \citet{Okamoto+2008}.  The dotted line
  represents the analytical solution for $M_F$ which is only valid before
  reionization for a power-law temperature evolution.}
\label{fig:M_F}
\end{figure*}

The associated mass scale, defined in analogy with the Jeans mass, is
\begin{equation}
  \label{eq:M_F}
  M_F(a) \equiv \frac{4\pi}{3}\,\bar{\rho}(a)\,\left(\frac{2\pi a}{k_F(a)}\right)^3\,,
\end{equation}
(the details of how this is computed in practice are collected in Appendices
\ref{sec:MF} and \ref{sec:MF_EdS}).  To understand the physics underlying the
filtering mass, it is instructive to connect the filtering mass to the entropy
of the IGM. To this end, we use the definition for the entropy $K$ of
Equation~\eqref{eq:Ke}. As we show in Appendix \ref{sec:MF},  the filtering scale $\lambda_F
= 2\pi a/k_F$ is an integral over the entropy evolution,
\begin{eqnarray}
  \label{eq:k_F_entropy}
  \frac{1}{k_F^2(a)} &=& \frac{ A_0}{D_+(a)}\int_0^a da' K(a')
  \frac{D_+(a')}{a'^3 E(a')} \int_{a'}^a  \frac{da''}{a''^3 E(a'')}, \\
  A_0 &=& \frac{5}{3}\,\left(\frac{3\,\Omega_m}{8\pi G H_0}\right)^{2/3}.
\end{eqnarray}
Here $E(a) = H(a)/H_0$ is the dimensionless Hubble function. Since $M_F\propto
k_F^{-3}$, the linear filtering mass is also determined by the entropy evolution
at mean density and appropriately scaled with the linear growth function, Hubble
function, and cosmic scale factor.  Since the integrand in
Equation~(\ref{eq:k_F_entropy}) is a positive definite function, the linear
filtering mass increases monotonically with time.

We expect the strong increase of the entropy due to blazar heating to drive a
comparably strong increase in $M_F$, though slightly delayed due to the
effective weighting function in $M_F$. Physically, this implies that the entropy
floor delivered by photo- or blazar heating at densities around the cosmic mean
is then conserved by adiabatic compression during structure formation and
directly translates to a linear filtering mass $M_F$. However, during this
nonlinear process of structure formation, the entropy can be additionally
augmented through dissipation of gravitational energy in cosmic
structure-formation shocks or decreased through radiative cooling. The
competition of these two processes determines which will dominate and in turn
sets the critical mass $M_C$ which is required for the condensation of gas into
a dark matter halo and eventually the formation of a galaxy.

As with $M_J$, $M_F$ typically exceeds the $M_C$ observed in simulations by a
substantial factor.  Whether a halo can accrete gas is determined by the gas
temperature (or equivalently entropy) at the virial radius which can only be
obtained through hydrodynamic cosmological simulations.  Estimated values for
this critical mass using cosmological simulations of nonlinear structure
formation with only photoheating yield $M_C (z=0) = 6.5\times 10^9 h^{-1}
M_\sun$ \citep{Hoeft+2006,Okamoto+2008}---one order of magnitude smaller than
the linear analog $M_F$ at $z=0$ \citep[][see also
Figure~\ref{fig:M_F}]{Okamoto+2008}. This discrepancy becomes less for
increasing redshift leading to reasonably good agreement at high redshift ($z >
6$) between $M_F$ and $M_C$. To model the nonlinear behavior of $M_F$, we can
introduce a correction factor $C(z) = M_C(z) / M_F(z) = (1+z)^{1.1}/11.8$, where
$M_F(z)$ has been modeled from the temperature evolution at mean density in the
cosmological simulations and includes weakly nonlinear aspects of structure
formation \citep{Okamoto+2008, Maccio+2010}.  However, validating this
approximation for $M_C$ will ultimately require numerical simulations that
include blazar heating.

Figure \ref{fig:M_F} shows $M_F$ as a function of redshift for a void and the
cosmic mean when various heating mechanisms are considered \citep[which agrees
for the standard cosmology with Figure 5 of][]{Gnedin2000}.  Note that for all
of the thermal histories we considered, the resulting $M_F$ nicely follows the
expected analytical form over its full range of validity prior to recombination
(which necessarily implies an Einstein-de Sitter universe, see the derivation in
Appendix \ref{sec:MF_EdS}). While blazar heating increases $M_F$ in voids as
early as $z\sim3$, perturbations at the cosmic mean are affected slightly later
($z\sim2.5$).  The extra suppression of the formation of small galaxies due to
blazar heating amounts to a mass suppression factor of $\sim25-70$ for voids and
$\sim6.5-15$ for the cosmic mean today. The range indicates the uncertainty of the
blazar heating models and the larger value belongs to the model that matches the
inverted temperature-density relation in the \Lya forest found by \citet{Viel+09}, i.e.,
the optimistic model.  The resulting estimates for the nonlinear characteristic
mass $M_C$ are also shown in Figure~\ref{fig:M_F} with red lines for our models
with and without blazar heating. Apparently the increase of $M_C$ due to blazar
heating is able to counteract the suppression of $M_F$ in nonlinear theory. We
note that our estimate for $M_C$ is slightly decreasing for $z<1$ in the model
that employs only photoheating. This decrease is an artifact of our linear
theory $M_F$ which by construction does not model nonlinear aspects of
structure formation such as shell crossing or formation shocks that raise the
temperature of some patches already at mean density.

We now turn to two outstanding problems in cosmological structure formation that
the recent blazar heating may help to address: the missing satellite (Section
\ref{sec:MWMS}) and void dwarf (Section \ref{sec:VD}) problems.  Finally in
Section \ref{sec:WDM} we discuss how blazar heating may naturally address some
of the difficulties with WDM cosmologies of galaxy formation.

\subsection{Dwarf satellites in the Milky Way} \label{sec:MWMS}

The heating due to blazars provides an additional mechanism to suppress the
formation of dwarfs.  Unlike photoionization models, which typically invoke the
heating at reionization, blazar heating provides a well defined, time-dependent
suppression mechanism, with the suppression rising dramatically after $z\sim2$.
This can be seen explicitly by the steep increase in filtering mass for these
redshifts in Figure~\ref{fig:M_F}.  In addition, due to the homogeneous nature
of the heating, i.e., a constant volumetric rate that is independent of density,
the heating from blazars suppresses structure formation most efficiently in the
low-density regions responsible for late-forming dwarf halos due to their
negative bias, where the energy deposited per baryon is larger.

In fact, the star formation histories of dwarf galaxies provide a strong
constraint upon the magnitude of blazar heating: if dwarf galaxy formation is
suppressed by high-energy gamma-ray emission from blazars, all dwarf star
formation histories must begin prior to $z\simeq2$, roughly the redshift at
which blazar heating becomes significant.  Over the past decade most of Local
Group dwarfs have been observed with the {\em Hubble Space Telescope} shifting
the focus from counting dwarfs to resolving the individual stellar population
within these objects.  Thus, it has become possible to construct detailed
color-magnitude diagrams of every dwarf galaxy and therefore their detailed star
formation histories \citep{Dolphin+2005,Holtzman+2006,Oban+2008}.  While the
data shows a great variety of star formation histories---some continuous, some
bursty, some truncated---they all have in common that they  extend beyond a
lookback time of $10\,\Gyr$ (corresponding to $z=2$).  There is no known case of
a dwarf galaxy that formed its first set of stars after $z\simeq2$.  The fact that
there exists an old stellar population in every dwarf in the Local Group
provides a very important test that our model of blazar heating successfully
passes.

The degree to which blazar heating can suppress the number of dwarf galaxies
depends upon the redshifts at which they are typically formed.  By combining
high-resolution $N$-body simulations of the evolution of Galaxy-sized halo with
semianalytic models of galaxy formation, \citet{Maccio_WDM2010} and
\citet{Maccio+2010} have inferred the statistics, formation
time,\footnote{\citet{Maccio+2010} defined the formation time of dwarfs as the
  redshift where the progenitors of dwarfs that end up within the Milky Way halo
  today exceeded a virial temperature of $T_{200}=10^4\,\K$ such that H~{\sc i}
  cooling of the gas becomes possible in these halos. That virial temperature
  corresponds to a mass threshold of $M_{200}\simeq 10^9\,M_\sun$ at $z=1$ and
  $10^8\,M_\sun$ at $z=10$.} and accretion histories of the Milky Way
satellites.  Their findings are in good agreement with recent work by other
groups \citep{Koposov+2009, Munoz+2009, Guo+2010, Busha+2010, Font+2011} and
thus can be regarded as representative for this model class.  They find that the
distribution of formation times is bimodal as a result of the suppression of hot
gas accretion in low-mass halos due to the photoionization background; while
$2/3$ of today's satellites form at redshifts ranging from $3<z<12$, $1/3$ of
all satellites form late at $z < 3$ with most of them at $z < 1.5$.  Because
their formation is marked by the first time H~{\sc i} cooling can effectively
form the first stars, their stellar populations are necessarily younger than
their formation time of $z<1.5$ in direct conflict with observed ages of the
stellar populations in all of the Local Group dwarfs\footnote{This assumes star
  formation in primordial gas. However, other processes such as metal pollution
  from adjacent galaxies might cause stars to form earlier at lower virial
  temperatures of $T_{200}<10^4\,\K$.}
\citep{Dolphin+2005,Holtzman+2006,Oban+2008}.  The model of
\citet{Maccio_WDM2010} that successfully reproduced the satellite luminosity
function of the Milky Way assumed the linear theory filtering mass \citep[][see
solid blue line in Figure~\ref{fig:M_F}]{Gnedin2000} which was shown to
significantly overproduce the characteristic halo mass scale below which baryons
cannot condense and form stars \citep[][see solid red line in
Figure~\ref{fig:M_F}]{Okamoto+2008}. Adopting their less efficient nonlinear
filtering mass formalism, the number of satellite galaxies with intermediate
magnitudes $M_V \sim 10$ increases and creates a bump in the luminosity function
which is clearly inconsistent with the data \citep{Maccio+2010}.

Inspection of nonlinear estimates for the filtering mass $M_F$ in Figure
\ref{fig:M_F} shows that heating from blazars could prevent the condensation of
baryons in halos of masses $10^{10} M_\sun$ at redshifts $z\sim 2$ and likely up
to $10^{11} M_\sun$ at $z=0$.  At mean density, these nonlinear estimates for
$M_F$ correspond to the original linear values of \citet{Gnedin2000} which have
been shown to yield the observed dwarf satellite abundances and luminosity
functions \citep{Somerville2002,Maccio+2010}.  At the same time, the strongly
rising $M_F$ in the blazar heating models after $z\sim 2$ is able to suppress
the population of late-forming dwarfs \citep{Maccio_WDM2010} that are in
conflict with measured star formation histories of dwarf galaxies.  Thus, not
only can blazar heating potentially play a significant role in explaining the
observed abundances of dwarf galaxies but also their old star formation
histories.

Finally, we argue that the redshift evolution of the blazar heating rate should
manifest itself as stochasticity in the satellite luminosity function at fixed
host halo mass. Host halos show a distribution of formation times
\citep[e.g.,][]{Wechsler+2002}. This distribution is inherited by the host's
satellite dwarfs which on average form earlier than the host. As demonstrated in
Figure~\ref{f:thermal history}, blazar heating implies an entropy floor that
dramatically increases after $z\sim2$.  Hence, the distribution of formation
times of satellite dwarfs at fixed host halo mass results in a distribution of
pre-collapsed entropy of these dwarf halos. As a result, this leaves
early-forming dwarfs relatively unchanged but suppresses the baryon fraction or
even the formation of late-forming dwarfs. This results in different cooling
histories for different dwarfs, which, in turn, might modify the stellar content
that condenses out in these systems.  Hence, blazar heating provides an
apparently substantial, physically motivated scatter in the satellite luminosity
function, or distribution of mass-to-light ratio at fixed host halo mass. This
physically motivated stochasticity has important implications for abundance
matching techniques that need to be taken into account and complicates their
use. We note that some of this stochasticity is already seen in mildly nonlinear
theory (which captures some aspects of the formation history) and manifests
itself as a scatter in entropy at large 1+$\delta$ in Figure~\ref{fig:entropy
  blazar}.

\subsection{The Void Phenomenon and the Faint-End Slope of the Galaxy Luminosity
  and H~{\sc i}-Mass Function} \label{sec:VD}

As outlined in Section~\ref{sec:intro_dwarf}, the ``void phenomenon'' is closely
related to the substructure problem. Both show a strong discrepancy in the
abundance of dark matter (sub-)halos and paucity of luminous dwarf galaxies that
are thought to be hosted by these halos. Before we discuss in detail how blazar
heating impacts on dwarf galaxies in voids, we review the latest discussion on
the status of the problem itself which has recently been disputed to even exist.

Using a halo occupation distribution model approach, \citet{Tinker+2009} claim
to have explained the problem as they find agreement between luminosity
functions, nearest neighbor statistics, and void probability function of faint
galaxies.  However, a very high resolution simulation of the local volume of 8
Mpc around the Milky Way predicts a factor of 10 more dwarf halos than observed
dwarf galaxies in mini-voids with sizes ranging from 1 to 4.5 Mpc, hence
reinforcing the ``void phenomenon'' \citep{TikhonovKlypin2009}. While the
agreement between theory and observations is good for dwarfs with masses
$M_{200m}\gtrsim10^{10}\,M_\sun$ (corresponding to maximum circular velocities
of $\vel_c\gtrsim 40~\rmn{km~s}^{-1}$),\footnote{We quote the values
  corresponding to the lower bounds in their models that assumed a normalization
  of the matter power spectrum of $\sigma_8=0.9$. This seems to be a
  conservative choice when comparing to the critical mass threshold of
  $M_{200m}>6\times10^{9}\,M_\sun$ or equivalently $\vel_c>35~\rmn{km~s}^{-1}$
  which assumed a $\sigma_8=0.75$. For consistency reasons with the literature,
  in Section~\ref{sec:structure formation}, this section, we define the virial mass,
  $M_{200m}$, as the mass of a sphere enclosing a mean density that is 200 times
  the {\em mean} density of the universe.} it fails below, suggesting that
\citet{Tinker+2009} did not sample the relevant mass scales.  Moreover, their
analysis only demonstrated the self-consistency of the halo occupation
distribution model and did not make a comparison between observations and the
predictions of $\Lambda$CDM.

The discrepancy of the luminosity function on dwarf scales is confirmed by other
recent studies that compare the circular velocity function of H~{\sc i} observations
with those obtained through dissipationless simulations \citep{Zavala+2009,
  Zwaan+2010, Trujillo-Gomez+2010}. At $\vel_c<80~\rmn{km~s}^{-1}$
(corresponding to $M_{200m}<10^{11}\,M_\sun$), the latest study finds a slight
deviation between theory and observations which amounts to a significant
overprediction of more than 10 times the number of observed systems at
$\vel_c=40~\rmn{km~s}^{-1}$. While completeness of the observed samples could be
an explanation of some of the differences, it is unlikely that it can account
for all of the observed effects since the blind H~{\sc i} sample of the HIPASS survey
is thought to be complete down to $M_{\rmn{H}\,\textsc{i}}<5.5\times10^{7}\,M_\sun$ out
to a distance of 5~Mpc \citep{Zwaan+2010}.  Since gas-rich galaxies dominate at
the low mass end of the luminosity function, their sample should give an
accurate measurement of the abundance of dwarfs if these galaxies contain enough
neutral gas to be detected.

While blazar heating significantly increases the entropy, and hence the
associated filtering mass, at mean density, it does so more dramatically within
the voids.  To address how this substantially increased heating manifests itself
in the number of void dwarfs, it is instructive to compare the formation
timescales of dwarf halos as a function of environment.  Void dwarf galaxies are
negatively biased and form later than their field or cluster analogs (halos of a
given mass tend to be older in clusters and younger in voids for masses smaller
than the typical mass scale that is presently entering the nonlinear regime of
perturbation growth). This is because for galaxies forming on a large-scale
underdense mode, more time has to elapse before these galaxies acquire enough
overdensity to decouple from the Hubble expansion and subsequently collapse.
Thus, the median redshift of formation for a dwarf galaxy with halo mass
$2.4\times 10^{10}\,\rmn{M}_\sun$ is $z_\rmn{form} = 2.1$ within cluster and 1.6
in voids \citep{Hahn2+2007}.  Moreover, the distribution of formation times of
void galaxies is more sharply confined around the median value than those for
clusters, which exhibit a long tail of formation redshifts, extending to
$z\sim6$ \citep{Hahn1+2007}.  Our estimate of $M_C$ in underdense regions with
$1+\delta=0.5$ demonstrates that blazar heating prevents the condensation of
baryons in halos of masses of $10^{10} M_\sun$ at redshifts $z<2.4$ and $2.8
\times 10^{11} M_\sun$ at $z=0$ (see Figure~\ref{fig:M_F}). In particular, halos
of $2 \times 10^{10} M_\sun$ can be suppressed after $z=1.8$ which is earlier
than the median formation redshift of these galaxies.  Depending on the exact
definition of voids, this suggests that more than half of the galaxies with
masses $M<2 \times 10^{10} M_\sun$, corresponding to a maximum circular velocity
of $\vel_c<45~\rmn{km\,s}^{-1}$, can be suppressed or severely affected by blazar
heating. 

Such a preheated entropy floor not only suppresses dwarf formation at late
times ($z\lesssim2$), but may also modify galaxy formation at the low-mass end.
Assuming that low-mass halos are embedded in a preheated medium with an entropy
floor of $10\,\rmn{keV\,cm}^2$ at $z\lesssim2$ simultaneously matches data at
the faint-end slope of the galaxy luminosity function as well as of the H~{\sc
  i}-mass function \citep{Mo+2005}. This heuristic assumption almost exactly
coincides with the predictions of our blazar heating models (see
Figure~\ref{f:thermal history}). As a result of such a preheating, only a
fraction of the gas in a proto-galaxy region would be able to cool and be
accreted into the final galaxy halo by the present time.  If the accreted gas
resides in the diffuse phase, it does not lose angular momentum to the dark
matter, thereby possibly continuing to form large galaxy discs in low-density
environments \citep{Mo+2002}.

In summary, the entropy floor and filtering mass due to blazar heating are
dramatically increased in voids as a result of the constant volumetric heating
rate, which leads to an inverted temperature-density relation in low-density regions
(Paper II).  In combination with the later formation epoch of dwarfs at these
low densities, this implies a very efficient mechanism for suppressing void
dwarf formation in collapsed dark matter halos. Hence, our model provides an
elegant physical solution to the void phenomenon described by
\citet{Peebles2001}.

\subsection{Suppression of Dwarfs in Warm Dark Matter Cosmologies} \label{sec:WDM}

Recent dissipationless $\Lambda$CDM simulations produce not only far too many
dark matter satellites but also show that the most massive subhalos in
simulations of the Milky Way may be too dense to host any of its observed bright
satellites with luminosities $L_V > 10^5 \rmn{L}_\sun$, i.e., these massive
satellites in simulations attain their maximum circular velocity at too small
radii in comparison to the observed dwarf satellites
\citep{Boylan-Kolchin+2011,Boylan-Kolchin+2012}. These dark subhalos have
circular velocities at infall of $30-70\,\rmn{km~s}^{-1}$ and infall masses of
$(0.2-4) \times 10^{10} \rmn{M}_\sun$.  In principle, this puzzle can be solved
(or partially solved) in the following ways: by increasing the stochasticity of
galaxy formation on these scales, by reducing the central (dark matter)
densities by means of very efficient and violent baryonic feedback processes
acting on timescales much faster than the free-fall time, by assuming a total
mass of the Milky Way at the lower end of the allowed uncertainty interval
(i.e., $\sim8\times10^{11} \,\rmn{M}_\odot$) in combination with a shallower
subhalo density profile of the Einasto form as measured in simulations
\citep{Vera-Ciro+2012}, or by allowing these subhalos to initially form with
lower concentrations as would be the case, for example, if the dark matter were
made of warm, rather than cold particles \citep{Lovell+2011}.

In the limit of heavy WDM particles, e.g., the (sterile) neutrino with $m_\nu
c^2 \gtrsim 10\,\keV$ that is created in the early universe through mixing with
an active neutrino, structure formation proceeds almost indistinguishably from
CDM for all current observational probes \citep{Seljak+2006}.  For smaller
masses, however, the free streaming of neutrinos erases all fluctuations on
scales smaller than the free streaming length, which is roughly proportional to
their temperature and inversely proportional to their mass.  Using the \Lya
forest power spectrum measured by the Sloan Digital Sky Survey and
high-resolution spectroscopy observations in combination with CMB and galaxy
clustering constraints still allows for a neutrino with mass $m_\nu
c^2>2.5~\keV$ (95\% c.l.) that decoupled early while in thermal equilibrium
\citep{Seljak+2006}.\footnote{Since blazar heating dramatically changes the
  thermal history of the IGM (Paper II) and may be responsible for the inverted
  temperature-density relation at $z=2-3$ inferred by high-redshift \Lya studies
  \citep{Bolton+08,Viel+09,Puchwein+2011}, it is not clear whether these limits
  on sterile neutrino properties are weakened in the presence of blazar
  heating.}

Dissipationless high-resolution simulations of the evolution of a Galaxy-sized
halo have shown that (sub-)halos form and are accreted later onto the main halo
in WDM scenarios compared to the standard CDM paradigm, due to the lack of power
on small scales in WDM \citep{Maccio_WDM2010}.  In WDM scenarios with relatively
low values of the particle masses, $m_\nu c^2 = (2-5)$~keV, there are almost no
halos with $z_\rmn{form} \geq 11$, suggesting that the fraction of late-forming
dwarfs $z_\rmn{form} \lesssim 1.5$ is increased over the CDM scenario. This late
formation epoch of dwarfs reinforces the star formation history problem of Local
Group dwarfs: these late-forming dwarfs contain young stellar populations, in
direct conflict with the old ages of the stellar populations ($\tau>10$~Gyr) in
all of the Local Group dwarfs \citep{Dolphin+2005,Holtzman+2006,Oban+2008}.

However, these are precisely the objects that blazar heating most strongly
affects, as the strongly rising filtering mass (or entropy floor) after $z\sim2$
is able to suppress the population of these late-forming dwarfs.  Hence, while
in this scenario the free streaming of WDM erases power on scales smaller than
dwarfs, blazar heating reconciles the theoretically expected and the observed
star formation histories and alleviates standard objections to galaxy formation
in WDM cosmologies.

\subsection{Impact on the formation of $L_*$ galaxies}

In the previous sections, we argued that a blazar-heated entropy floor is able
to suppress late dwarf formation in voids and the Milky Way as well as modify
the thermodynamical profiles of galaxy groups and clusters.  Hence, it is natural
to ask whether there would be any effect of blazar heating on $L_*$ galaxy
formation which represents the mass scale in between these two extremes.  Blazar
heating is not powerful enough to raise the mean temperature at $\delta=0$ of
even the most extreme patches above $10^5\,\K$ (see Figure~9 in Paper II). The
classical criteria of galaxy formation are a short cooling time compared to the
dynamical time and to the age of the universe, $t_\rmn{cool} \lesssim H^{-1}$
and $t_\rmn{cool} \lesssim t_\rmn{dyn}$
\citep{Rees-Ostriker1977,Silk1977,White-Rees1978}, which are easily fulfilled
even in the presence of blazar heating \citep[see, e.g., Figure 1
in][]{Rees-Ostriker1977}. This implies that these galaxies can radiate away the
additional entropy that the gas attained prior to collapse due to blazar heating
within a free-fall time. It is, however, interesting to speculate whether the
blazar-heated high-entropy gas at low redshift $z<1$ has any impact on the late
accretion of gas into the hot reservoir of baryons from which gas cools and
fuels the late time star formation. Higher entropy gas should shock further out
in the halo than pre-cooled gas of lower entropy which is denser and can provide
a larger ram pressure.  Blazar heated gas has high entropy and is more
dilute. Hence, it is more easily torqued (by dissipative processes or magnetic
fields) and therefore change its angular momentum distribution.  Thus, it is
implausible that blazar heating dramatically changes the ordinary mode of galaxy
formation, but we might anticipate blazar heating to starve the late-time
accretion and potentially slow down subsequent star formation. These ideas are
subject to verification by numerical simulations of galaxy formation.

\section{Conclusions}\label{sec:conclusions}

TeV blazar heating results in a dramatic increase in the entropy of the IGM
following He {\sc ii} reionization around $z\sim3.5$.  Since the IGM entropy evolution
is critical for the formation and structure of collapsed objects, blazars
heating has a significant impact upon both.  We have identified two mass ranges
(or classes of objects) for which the TeV blazar-induced entropy floor should
have a substantial effect. {\em Galaxy groups and clusters}, which are forming
near the peak entropy injection rate ($z\sim1$) and exhibit core entropies that
are comparable to that implied by blazar heating; and {\em dwarf galaxies},
which are susceptible to the rapidly rising entropy floor generated by blazars.
Below, we describe the consequences for each in more detail.

{\em Galaxy groups and clusters.} Immediately after formation, groups at fixed
mass should have a continuous distribution of core entropy values, depending on
the formation redshift and the temporally variable heating mechanism.  The fate
of these groups is determined by the ratio of the cooling time, $t_\rmn{cool}$,
to the timescale between cluster/group mergers, $t_\rmn{merger}$. If this ratio
is smaller than unity, the group can radiate the elevated core entropy away and
evolve into a CC which survives the successive hierarchical
growth. Alternatively, if $t_\rmn{cool}> t_\rmn{merger}$, merger shocks can
gravitationally reprocess the entropy cores and amplify them. Those groups can
then evolve into NCC systems.  Hence, it is not necessary to produce all of the
observed central entropy in the IGM before collapse, but it is also possible to
achieve this through gravitational heating, provided there is a certain minimum
entropy delivered by some putative heating process corresponding to a minimum
$t_\rmn{cool}$.  An increasing entropy floor also implies an increasing cooling
time, hence the cluster-averaged $t_\rmn{cool}/t_\rmn{merger}$ increases.  It
follows that systems that evolve into CC systems {\em today} are on average
early-forming, i.e., old systems. In contrast, NCCs are on average young
systems.

We argue that systems with a unimodal distribution in core entropy values after
group formation should naturally evolve into a bimodal distribution. The reason
for this are the two attractor solutions of the group/cluster system, driven by
the cooling instability and gravitational reprocessing of (temporally
increasing) elevated entropy cores, resulting in the observed populations of CC
and NCC systems, respectively. Such an elevated entropy core level in groups
might explain a population of X-ray dim groups with low gas fractions. We show
that the core entropy values in CC clusters have a very weak correlation with
the mechanical energy and power of X-ray cavities inflated by AGNs. Apparently,
$PdV$ work done by the expanding cavities is an inefficient heating process that
does not generate much entropy.  This strongly suggests that while AGN feedback
seems to be critical in stabilizing CC systems, it cannot transform CC into NCC
systems (at least on the buoyancy timescale).

Our blazar-induced entropy history seems to be well matched for the formation
times of today's groups, but at first sight less so for clusters which form in
highly biased regions through mergers of groups. Those had to form even earlier
when the entropy for the {\em average} IGM was still rising with typical values
at $z\simeq1$ of $K_0\simeq (25 - 50)\,\keV~\cm^2$.  However, two effects
positively interfere to counteract the apparently smaller effect of blazar
heating in clusters.  First, the mass accretion rate is larger for larger
systems and at higher redshifts.  This suggests that the earlier forming group
progenitors of clusters can tolerate a smaller blazar-heated entropy floor after
collapse which will then be gravitationally processed at a faster rate and able
to counteract the smaller cooling timescales.  Second, blazars also turn on
first in highly biased regions, and thus the IGM in the vicinity of clusters
should experience blazar heating earlier than low-density regions.  We speculate
that this effect in combination with the faster gravitational reprocessing of an
elevated entropy core in denser regions would help in propagating the effect of
blazar heating to the scale of massive clusters.

Changing the thermodynamic structure of groups and potentially clusters has also
an impact on the SZ power spectrum which is sensitive to cosmological parameters
such as $\sigma_8$ and the thermal pressure profile. An increased core entropy
level implies a smoother pressure profile and hence decreases the power in the
SZ power spectrum on scales smaller than $3\arcmin$ which are probed by current
experiments such as SPT and ACT. Lowering the astrophysical signal can allow for
larger values of $\sigma_8$, potentially reducing some (minor) tension with the
data especially after allowing for a contribution due to patchy reionization.

{\em Dwarf galaxy formation.} We demonstrate that the redshift-dependent entropy
floor increases the characteristic halo mass, $M_C$, below which dwarf galaxies
cannot form by a factor of 6.5--15 for the cosmic mean ($\delta=0$) and by a
factor of 25--70 for voids ($\delta=-0.5$). The range indicates the uncertainty
in the number of blazars that contribute to the heating and the upper envelope
matches the observations of an inverted temperature-density relation in the \Lya
forest. The increase of $M_C$ prevents the formation of late-forming dwarf
galaxies ($z\lesssim2$) with masses ranging from $10^{10}$ to
$10^{11}\,\rmn{M}_\sun$ for redshifts $z\sim2$ to 0, respectively.  This may
resolve the ``missing satellites problem'' in the Milky Way, i.e., the low
observed abundance of dwarf satellites compared to CDM simulations. It also
brings the observed early star formation histories of Local Group dwarfs into
agreement with galaxy formation models that predicted a population of
late-forming objects, in conflict with the data.  At the same time, it provides
a plausible explanation for the ``void phenomenon'' which is the apparent
discrepancy of the number of dwarfs in low-density regions in CDM simulations
and the paucity of those in observations.  Blazar heating suppresses the
formation of galaxies within existing dwarf halos of masses $< 3\times
10^{10}\,\rmn{M}_\sun$ with a maximum circular velocity $< 60~\rmn{km~s}^{-1}$
for $z\lesssim2$. Additionally, the phenomenology of such a preheating mechanism
matches heuristic assumptions that were adopted to match the faint-end slope of
the galaxy luminosity function as well as of the H {\sc i}-mass function, in
particular for low-density environments.

We conclude that the presented scenario of blazar heating holds the promise for
solving some of the most outstanding problems in high-energy gamma-ray
astrophysics, the IGM as probed by the high-redshift \Lya forest, the formation
of galaxies, and clusters of galaxies. At the same time, it provides an
astrophysical solution to these problems such as the ``missing satellite
problem'' or the ``void phenomenon'' which have been claimed to require new
physics beyond the concordance cosmological $\Lambda$CDM model.

\acknowledgements We thank Tom Abel, Marco Ajello, Marcelo Alvarez, Arif Babul,
Roger Blandford, James Bolton, Mike Boylan-Kolchin, Luigi Costamante, Andrei
Gruzinov, Peter Goldreich, Martin Haehnelt, Andrey Kravtsov, Hojun Mo, Ue-li
Pen, Ewald Puchwein, Volker Springel, Chris Thompson, Matteo Viel, Marc Voit,
and Risa Wechsler for useful discussions.  We are indebted to Peng Oh for his
encouragement and useful suggestions. We thank Steve Furlanetto for kindly
providing technical expertise. We also thank the referee for a thorough reading
of the manuscript and for his constructive comments. These computations were
performed on the Sunnyvale cluster at CITA.  A.E.B. and P.C. are supported by
CITA. A.E.B. gratefully acknowledges the support of the Beatrice D. Tremaine
Fellowship.  C.P. gratefully acknowledges financial support of the Klaus Tschira
Foundation and would furthermore like to thank KITP for their hospitality during
the galaxy cluster workshop.  This research was supported in part by the
National Science Foundation under Grant No. NSF PHY05-51164.

\appendix

\section{Computing the Filtering Mass Generally}\label{sec:MF}
Here we collect the relevant details for computing the filtering mass,
$M_F$, associated with a particular cosmological and thermodynamic
evolution of the IGM.  Recall that the filtering wavenumber is related
to the Jeans wavenumber via
\begin{multline*}
  \frac{1}{k_F^2(t)} = \frac{1}{D_+(t)}\int_0^t dt' a^2(t') 
  \frac{\ddot{D}_+(t') + 2 H(t')\dot{D}_+(t')}{k_J^2(t')}\\
  \times\int_{t'}^t \frac{dt''}{a^2(t'')}\,.
\tag{\ref{eq:Jeans_Dplus}}
\end{multline*}
We can simplify Equation~(\ref{eq:Jeans_Dplus}) by noting that the
linear growth function obeys the following equation
\begin{equation}
  \label{eq:Dplus1}
  \ddot{D}_+(t) + 2 H(t)\dot{D}_+(t) = 4\pi G \bar{\rho}D_+(t),,
\end{equation}
in which the Hubble function is given by
\begin{equation}
  \label{eq:Hubblea}
  \frac{H^2(a)}{H_0^2} = E^2(a) = 
  a^{-3} \Omega_m + a^{-2} (1-\Omega_m - \Omega_\Lambda) + \Omega_\Lambda\,.
\end{equation}
Substituting the integration variable in Equation~(\ref{eq:Jeans_Dplus}) by the
scale factor $a$, we obtain
\begin{equation}
  \label{eq:k_F}
  \frac{1}{k_F^2(a)} = \frac{1}{D_+(a)}\int_0^a da' \frac{c_s^2(a')}{H_0^2}
  \frac{D_+(a')}{a'E(a')} \int_{a'}^a  \frac{da''}{a''^3 E(a'')}.
\end{equation}
Following \citet{1992ARA&A..30..499C}, the linear growth function can be
computed by
\begin{equation}
  \label{eq:Dplus2}
    D_+(a) = \frac{5}{2}\,\Omega_m \,E(a)\,\int_0^a da' \frac{1}{a'^3E^3(a')}.
\end{equation}
For any given temperature evolution of the IGM (which enters via the sound
speed), we can compute a redshift evolution of the filtering scale from
Equation~(\ref{eq:k_F}). It is convenient to define a filtering mass by analogy
with the Jeans mass:
\begin{equation}
  M_F(a) \equiv \frac{4\pi}{3}\,\bar{\rho}(a)\,\left(\frac{2\pi a}{k_F(a)}\right)^3.
\label{eq:M_F_app}
\end{equation}
Finally, using a definition for the entropy of Equation~\eqref{eq:Ke}, we can
derive the filtering scale $\lambda_F = 2\pi a/k_F$ in terms of the entropy,
\begin{eqnarray}
  \label{eq:k_F_entropy_app}
  \frac{1}{k_F^2(a)} &=& \frac{ A_0}{D_+(a)}\int_0^a da' K(a')
  \frac{D_+(a')}{a'^3 E(a')} \int_{a'}^a  \frac{da''}{a''^3 E(a'')}, \\
  A_0 &=& \frac{5}{3}\,\left(\frac{3\,\Omega_m}{8\pi G H_0}\right)^{2/3}.
\end{eqnarray}

\section{Filtering mass in an Einstein-de Sitter universe}
\label{sec:MF_EdS}

For sufficiently early times, all matter-dominated Friedmann-Lema{\^i}tre model
universes asymptotically approach an Einstein-de Sitter universe ($z\gtrsim2$ for our
$\Lambda$CDM universe), for which the cosmological constant $\Lambda=0$ and the
curvature $\Omega_k=0$. For such a universe, the growth function $D_+(a) \propto a$
which allows us to considerably simplify Equation~(\ref{eq:k_F}), yielding
\begin{equation}
  \label{eq:kF_EdS}
  \frac{1}{k_F^2(a)} = \frac{3}{a} \int_{a_\mathrm{min}}^a da'\,\frac{1}{k_J^2(a')}\,
  \left[1-\left(\frac{a'}{a}\right)^{1/2}\right].
\end{equation}
Note that we replaced the unphysical lower integration limit 0 by the recombination
scale factor, $a_{a_\mathrm{min}}$, as baryon perturbations can only start to
grow after recombination.  Employing the relation of Equation~(\ref{eq:M_F_app}) we can
rewrite this to give
\begin{equation}
  \label{eq:MF_EdS}
  M_F^{2/3}(a) = \frac{3}{a} \int_{a_\mathrm{min}}^a da'\,M_J^{2/3}(a')\,
  \left[1-\left(\frac{a'}{a}\right)^{1/2}\right].
\end{equation}

After recombination, the residual electron density couples the gas temperature
still to that of the CMB via Compton interactions. Hence, we expect the gas
temperature to scale as $T\propto a^{-1}$.  At $z\simeq 150$, the Compton
interaction rate drops below the Hubble expansion rate such that the gas
experiences adiabatic expansion with $T\propto a^{-2}$. Hence for the time after
matter-radiation equality, where $\bar{\rho}=\bar{\rho}_{0}\,a^{-3}$, we can write
in general
\begin{equation}
  \label{eq:PL_temp}
  \frac{T(a)}{T_0} = \left(\frac{a}{a_0}\right)^{-\alpha},\quad\mbox{and}\quad
  \frac{k_J(a)}{k_{J,0}} = \left(\frac{a}{a_0}\right)^{(\alpha-1)/2}. 
\end{equation}
This definition for the Jeans scale implies a Jeans mass at a fiducial scale
factor, $a_0$, of
\begin{equation}
  \label{eq:MJ0}
  M_{J,0} \equiv \frac{4\pi}{3}\,\bar{\rho}(a_0)\,\left(\frac{2\pi a_0}{k_{J,0}}\right)^3 .
\end{equation}
If we substitute Equation~(\ref{eq:PL_temp}) into Equation~(\ref{eq:MF_EdS}), we can obtain an
analytical solution for the filtering mass before reionization,
\begin{equation}
  \label{eq:MF_EdS_alpha}
  M_F(a) = M_{J,0} \left\{
      \frac{3}{a} \int_{a_\mathrm{min}}^a da'\,\left(\frac{a}{a_0}\right)^{1-\alpha}\,
      \left[1-\left(\frac{a'}{a}\right)^{1/2}\right]\right\}^{3/2}.
\end{equation}
Evaluating this integral for the two regimes before and after the freezeout of
Compton interactions of the gas with the CMB photons, yields the following analytic solutions,
\begin{equation}
M_F(a) 
=
M_{J,0} \left\{ 
\begin{aligned}
&\left[\frac{6a_0}{a}\left(\sqrt{\frac{a_\mathrm{min}}{a}} - 1 
+ \frac{1}{2}\,\ln \frac{a}{a_\mathrm{min}}\right)\right]^{3/2}\\
&\qquad\qquad\qquad\qquad{\rm for}~150<z<1100\\
&\left[1 - 3\,\frac{a_{\mathrm{min}}}{a} +
  2\,\left(\frac{a_{\mathrm{min}}}{a}\right)^{3/2}\right]^{3/2}\\
&\qquad\qquad\qquad\qquad{\rm for}~z_\rmn{reion}\le z\le150\,.
\end{aligned}
\right.
\label{eq:MF_high-z}
\end{equation}

\bibliographystyle{apj}

\end{document}